\documentclass[fleqn,usenatbib]{mnras}
\usepackage{newtxtext,newtxmath}
\usepackage[T1]{fontenc}

\pdfoutput=1
\usepackage{aas_macros}
\usepackage{epsfig}
\usepackage{amsmath}
\usepackage{amssymb}
\usepackage{wasysym}

\usepackage{fancyref}
\usepackage{hyperref}
\hypersetup{colorlinks=true,citecolor=cyan}

\def\beq{\begin{equation}}
\def\eeq{\end{equation}}
\def\ba{\begin{eqnarray}}
\def\ea{\end{eqnarray}}
\def\bal{\begin{align}}
\def\eal{\end{align}}
\def\bxi{{\mbox{\boldmath $\xi$}}}

\title[Tidally Trapped Pulsations]{Tidally Trapped Pulsations in Binary Stars}

\author[J. Fuller et al.]{
J. Fuller,$^{1}$\thanks{E-mail: jfuller@caltech.edu}
D. W. Kurtz,$^{2,3}$
G. Handler,$^{4}$
S. Rappaport$^{5}$
\\
$^{1}$TAPIR, Mailcode 350-17, California Institute of Technology, Pasadena, CA 91125, USA\\
$^{2}$Centre for Space Research, Physics Department, North West University, Mahikeng 2745, South Africa\\
$^{3}$Jeremiah Horrocks Institute, University of Central Lancashire, Preston PR1 2HE, UK\\
$^{4}$Nicolaus Copernicus Astronomical Center, Polish Academy of Sciences, ul. Bartycka 18, 00-716, Warszawa, Poland\\
$^{5}$Department of Physics, and Kavli Institute for Astrophysics and Space Research, M.I.T., Cambridge, MA 02139, USA
}

\date{Accepted XXX. Received YYY; in original form ZZZ}
\pubyear{2030}

\begin{document}
\label{firstpage}
\pagerange{\pageref{firstpage}--\pageref{lastpage}}
\maketitle

\begin{abstract}

A new class of pulsating binary stars was recently discovered, whose pulsation amplitudes are strongly modulated with orbital phase. Stars in close binaries are tidally distorted, so we examine how a star's tidally induced asphericity affects its oscillation mode frequencies and eigenfunctions. We explain the pulsation amplitude modulation via tidal mode coupling such that the pulsations are effectively confined to certain regions of the star, e.g., the tidal pole or the tidal equator. In addition to a rigorous mathematical formalism to compute this coupling, we provide a more intuitive semi-analytic description of the process. We discuss three resulting effects: 1. Tidal alignment, i.e., the alignment of oscillation modes about the tidal axis rather than the rotation axis; 2. Tidal trapping, e.g., the confinement of oscillations near the tidal poles or the tidal equator; 3. Tidal amplification, i.e., increased flux perturbations near the tidal poles where acoustic modes can propagate closer to the surface of the star. Together, these phenomena can account for the pulsation amplitude and phase modulation of the recently discovered class of ``tidally tilted pulsators.'' We compare our theory to the three tidally tilted pulsators HD~74423, CO~Cam, and TIC~63328020,  finding that tidally trapped modes that are axisymmetric about the tidal axis can largely explain the first two, while a non-axisymmetric tidally aligned mode is present in the latter. Finally, we discuss implications and limitations of the theory, and we make predictions for the many new tidally tilted pulsators likely to be discovered in the near future.


\end{abstract}


\section{Introduction}
\label{intro}

Astrophysicists are fortunate that most stars are nearly spherical, allowing for straightforward computations of stellar pulsation modes in terms of spherical harmonics. Stars in very close binary systems, however, are heavily tidally distorted and very aspherical. It follows that the oscillation modes of such stars could be greatly affected, and the purpose of this paper is to examine the formalism and phenomenology of pulsating tidally distorted stars. 

Several prior works have investigated the effects of tidal distortion on stellar oscillation modes, focusing largely on the effect of tidal frequency splitting. If one ignores any effects of rotation, a tidally distorted star is symmetric about the tidal axis, i.e., the line of apsides joining the tidally distorted star with its companion. The axis of symmetry for the oscillation modes then becomes the tidal axis, and a mode multiplet of angular degree $\ell$ is split into $\ell+1$ peaks for each value of $|m|$, where $m$ is the azimuthal wave number about the tidal axis.
Several authors have investigated this process at increasing levels of complexity \citep{chandrasekhar1963,chandrasekhar1963b,tassoul1967,denis1972,saio1981,martens1982,smeyers1983,martens1986,reyniers2003,reyniers2003b}. 

Using a WKB ray tracing approach, \cite{springer2013} showed how acoustic waves could be focused onto different sides of the star, altering their damping rates and visibility. As we shall see in this paper, this tidal focusing effect, which alters the mode eigenfunctions and visibility, is readily observable. However, the approach of \cite{springer2013} is inadequate for computing low-order (both in $\ell$ and radial order $n$) tidally modified oscillation modes like those that are typically observed, because their angular length scale is similar to that of the tidal distortion, and a WKB approach not suitable.

The observational manifestation of pulsations aligned with the tidal axis are similar to the magnetic Ap star oblique pulsators (\citealt{1982MNRAS.200..807K}, \citealt{1985PASJ...37..245S},  \citealt{1993PASJ...45..617S}, \citealt{1995PASJ...47..219T}, \citealt{2002A&A...391..235B}, \citealt{Saio2004}, \citealt{2011A&A...536A..73B}), whose pulsations are primarily aligned with the magnetic axis due to Lorentz forces. Pulsations of tidally aligned pulsators would then be modulated over the orbital phase due to the changing viewing angle relative to the tidal axis.

If one assumes that modes of different $\ell$ remain uncoupled, then a mode of angular degree $\ell$ viewed in the corotating frame would be split into up to $2 \ell +1$ frequencies (separated by the orbital frequency) in the observer's frame \citep{Balona2018}. This is in addition to the tidal splitting of each mutliplet into $\ell+1$ components in the frame corotating with the orbit, so the observed pulsation spectrum of such a star could be very complex. As we shall show in this work, coupling between modes of different $\ell$ is very important, further altering the expected signal. Due in part to this complexity, there were few (if any) examples of tidal modulation of stellar pulsations until very recently.

\cite{handler2020} presented the first clear discovery of tidally trapped pulsations in the HD~74423 binary system, containing a $\delta$~Sct pulsator in a $\simeq \! 1.58$-d orbital period binary. The amplitude and phase modulation of the single oscillation mode detected in that system, a low-order p~mode at a frequency of $8.8 \, {\rm d}^{-1}$, made it clear that the mode was not only aligned with the tidal axis, but that it was strongly confined to one side of the star facing either the first or third Lagrange points (${\rm L}_1$ or ${\rm L}_3$). CO~Cam \citep{kurtz2020} was the second discovered system, with the notable difference that it showed at least four separate tidally trapped oscillation modes with frequencies near $14 \, {\rm d}^{-1}$, each trapped near the ${\rm L}_1$ point. The recently discovered system TIC~63328020 exhibits a tidally tilted pulsation at $21 \, {\rm d}^{-1}$,  but in this case the mode is non-axisymmetric about the tidal axis and is consistent with a distorted sectoral dipole mode of $\ell =1, |m|=1$. Finally, the U Gru system \citep{bowman2019} may also exhibit tidally trapped pulsations, but the nature of the complex behavior in that system remains unclear. 

We note that tidal trapping is a totally distinct phenomenon from tidal excitation of pulsations, like tidally forced gravity modes seen in heartbeat stars \citep{Fuller_2017}. Tidal excitation stems from the time-varying tidal force (i.e., the dynamical tide) exerted in an eccentric or non-synchronized binary that forces modes to oscillate at exact integer multiples of the orbital frequency. In contrast, tidal trapping arises from the static tidal distortion (i.e., the equilibrium tidal response) of a star, altering the shape of the pulsation cavity and hence the character of the star's free oscillation modes.

In this paper, we investigate the physical effect of tidal distortion on a pulsating star. We consider the simplest case of a tidally synchronized pulsating star in a circular orbit, such that the tidal distortion is time-independent in the corotating frame of the pulsator. Unlike many previous investigations, which focused on the tidal splitting of a single mode multiplet, we focus on tidal coupling between multiplets with different angular number $\ell$ and radial order $n$ (but the same azimuthal number $m$ about the tidal axis). This coupling allows for tidal trapping because the mode eigenfunctions become a superposition of spherical harmonics with different values of $\ell$, such that mode displacements and flux perturbations can have larger values on one side of the star, as is now clearly observed in several systems. 

We begin by presenting a mathematical formalism to account for tidal distortion and coupling between eigenmodes, for use in detailed calculations. We also provide some analytic and heuristic arguments to build intuition for how this coupling can lead to tidally trapped pulsations. We apply our detailed calculations to models of HD~74423, CO~Cam, and TIC~63328020, showing that the models very naturally predict tidally trapped pulsations whose amplitude modulation closely matches that which is observed. Finally, we discuss our results and make predictions for future observations.

\section{Tidal Mode Coupling}

\subsection{Mathematical Formalism}
\label{formalism}

In most stars, the dominant force breaking the spherical symmetry is caused by rotation: Coriolis forces are typically most important for low-frequency g~modes, while the centrifugal distortion of the star becomes increasingly important for high-frequency p~modes. These forces not only cause frequency splitting of mode multiplets, but they also cause coupling between other multiplets with different values of $\ell$ and $n$. In both cases, the star's symmetry axis is the rotation axis, so this is almost always assumed to be the pulsation axis. In a non-rotating tidally distorted star, however, the symmetry axis is the tidal axis (i.e, the line of apsides connecting the star and its companion), so the pulsation axis will realign accordingly. 

We consider the case of a tidally distorted star of mass $M_1$ in a binary system with a circular orbit of period $P$ and orbital separation $a$, whose spin is aligned and synchronized with the orbital motion of its companion, with mass $M_2$. In the rotating frame of the star, the tidal distortion is static, so we work in this frame. In this paper, for simplicity we will largely ignore the centrifugal distortion and Coriolis forces and focus on tidal distortion. We do include components of the centrifugal force that are axisymmetric about the tidal axis, and we discuss limitations of this approximation in Section \ref{disc}. Neglecting rotation, the perturbed star is symmetric about the tidal axis, and we work in spherical coordinates about this axis. We refer to the colatitude $\theta$ as the tidal latitude to distinguish it from the colatitude measured from the rotation axis. The tidal latitude is defined to have $\theta = 0$ in the direction of the companion star. In this frame, the tidal potential $U$ only has components with azimuthal number $m=0$, and can be expanded in spherical harmonics of angular degree $\ell_t$ as
\beq
\label{Utide}
U = \sum_{\ell_t} U_{\ell_t} = - \frac{G M_2}{a} \sum_{\ell_t=2}^\infty \sqrt{\frac{4 \pi}{2 \ell_t +1}} \bigg( \frac{r}{a} \bigg)^{\! \ell_t} Y_{\ell_t 0} (\theta, \phi) \, ,
\eeq
where $r$ is the radial coordinate from the center of $M_1$.

The tidal force distorts the shape of the star, meaning that the differential operators that describe pulsation modes in a spherical star must be mapped into a new coordinate system of the tidally distorted star. This mapping process was described in great detail by \cite{saio1981,martens1982,reyniers2003,reyniers2003b}. The mapping can be described using perturbation theory by considering small changes to the operators that define eigenmodes of the system. We follow the procedure and notation developed for terrestrial seismology in \cite{dahlen1998}, which we find to be easier to follow and more comprehensive than most astrophysical references. This procedure was also used for Coriolis and centrifugal forces in \cite{fullersatb:14}.

For a spherical star, the eigenmodes are defined as solutions of an equation
\beq
\label{eigen}
\mathcal{V} \bxi_\alpha = \omega_\alpha^2 \mathcal{T} \bxi_\alpha \, .
\eeq
Here, $\bxi_\alpha$ is the displacement vector of an eigenmode with eigenfrequency $\omega_\alpha$, $\mathcal{T}$ is the kinetic energy operator, and $\mathcal{V}$ is a potential energy operator that accounts for relevant forces (pressure, buoyancy, etc.). Equation \ref{eigen} demonstrates the usual equipartition of potential and kinetic energy for stellar oscillation modes. The eigenmodes are orthonormal such that the overlap between two eigenmodes is
\begin{align}
\label{Tab}
    \mathcal{T}_{\alpha \beta} &= \langle \xi_\alpha \vert \mathcal{T} \vert \xi_\beta \rangle \nonumber \\
    & = \int dV \rho \bxi_\alpha^* \cdot \bxi_\beta \, \nonumber \\
    & = \delta_{\alpha \beta} \, 
\end{align}
with $\delta_{\alpha \beta}$ the Kronecker delta. Additionally,
\begin{align}
\label{Vab}
    \mathcal{V}_{\alpha \beta} &= \langle \xi_\alpha \vert \mathcal{V} \vert \xi_\beta \rangle \nonumber \\
    & = \int dV \bxi_\alpha^* \cdot \big( \nabla \delta P_\beta + \delta \rho_\beta {\bf g} + \rho \delta {\bf g}_\beta \big)  \, \nonumber \\
    & = \omega_\alpha^2 \delta_{\alpha \beta} \, 
\end{align}
and $\delta P$, $\delta \rho$, and $\delta {\bf g}$ are the Eulerian perturbations to pressure, density and gravity. 

In a tidally distorted star, the modes satisfy the new equation
\beq
\label{TVpert}
\big[\mathcal{V} + \delta \mathcal{V} \big] \bxi = \omega^2 \big[\mathcal{T} + \delta \mathcal{T} \big] \bxi \, 
\eeq
where $\delta \mathcal{T}$ and $\delta \mathcal{V}$ are the changes to the kinetic and potential energy operators due to the tidal force and the aspherical geometry. We solve equation \ref{TVpert} by expanding the eigenvectors $\bxi$ in terms of the star's unperturbed set of modes $\bxi_\alpha$, so that
\beq
\bxi = \sum_\alpha a_\alpha \bxi_\alpha \, ,
\eeq
and $a_\alpha$ is the projection onto each original mode. Plugging this into equation \ref{TVpert},
\beq
\label{TVnew}
\big[\mathcal{V} + \delta \mathcal{V} \big] \sum_\alpha a_\alpha \bxi_\alpha = \omega^2 \big[\mathcal{T} + \delta \mathcal{T} \big] \sum_\alpha a_\alpha \bxi_\alpha \, .
\eeq
Multiplying by an arbitrary mode $\bxi_\beta$ and integrating over volume, and using equations \ref{Tab} and \ref{Vab}, we find 
\beq
\label{sum}
\sum_\alpha (\omega_{\alpha}^2 \delta_{\alpha \beta} + \delta \mathcal{V}_{\alpha \beta} ) a_\alpha = \omega^2 \sum_\alpha (\delta_{\alpha \beta} + \delta \mathcal{T}_{\alpha \beta} ) \, ,
\eeq
where $\delta \mathcal{T}_{\alpha \beta} = \langle \xi_\alpha \vert \delta \mathcal{T} \vert \xi_\beta \rangle$ and $\delta \mathcal{V}_{\alpha \beta} = \langle \xi_\alpha \vert \delta \mathcal{V} \vert \xi_\beta \rangle$. Equation \ref{sum} defines a matrix equation of the form 
\beq
\label{TV}
\mathbb{V} {\bf a} = \omega^2 \mathbb{T} {\bf a}
\eeq
that can be solved for the eigenfrequencies $\omega^2$ and eigenvectors ${\bf a}$, i.e., the expansion of the new eigenmodes in the basis of modes of the unperturbed star.

We provide explicit expressions for the matrix elements $\delta \mathcal{T}_{\alpha \beta}$ and $\delta \mathcal{V}_{\alpha \beta}$ in Appendix \ref{math}. They depend on the equilibrium tidal asphericity of the star, $\varepsilon_{\ell_t}$, for each component of the tidal potential. Because the perturbed modes are superpositions of unperturbed modes, the linearized equations and boundary conditions for the basis modes are unchanged, and only the overlap integrals in Appendix \ref{math} need to be computed.

To solve for the tidal distortion, we use the adiabatic and Cowling approximations, in which case 
the radial tidal displacement is \citep{goldreich1989}
\beq
\xi_{r, {\rm eq}} = -\frac{U}{g} \, ,
\eeq
with corresponding asphericity of $\varepsilon \sim \xi_{r,{\rm eq}}/r$ as defined in equation \ref{epsl}.
In practice, the neglect of the perturbed self-gravity is a very good approximation, as the size of this perturbation is $\approx k_2 U$, where the Love number $k_2 < 0.1$ for the early type stars that we will focus on. In this approximation, the Eulerian potential perturbation is simply $\delta \phi_{\rm eq} = U$, and the Lagrangian potential perturbation vanishes. However, we do {\it not} use the Cowling approximation when computing the unperturbed oscillation modes $\bxi_\alpha$ since we are interested in low-order modes (i.e., fundamental modes) for which self-gravity can be important. Our adiabatic approximation implies a vanishing Lagrangian perturbation of the pressure, density, and temperature due to the equilibrium tidal distortion. In reality, the perturbed radiative flux alters the entropy and temperature distribution along isobars, leading to gravity darkening \citep{vonZeipel:1924}. We expect these effects to produce modest quantitative changes in the tidal coupling coefficients, but not to qualitatively alter the results.

The tidal coupling coefficients depend on several angular overlap integrals that are a function of the indices $\ell$ and $m$ for each mode, as well as the $\ell_t$ of the tidal potential. In our setup, the tidal potential is axisymmetric and so only modes of $m_\alpha = m_\beta$ couple to each other. The angular integrals exhibit the usual three-mode coupling selection rule that only modes with $|\ell_\alpha + \ell_\beta| \geq \ell_t \geq |\ell_\alpha - \ell_\beta|$ and even values of $\ell_{\rm tide} + \ell_\alpha + \ell_\beta$ have a non-zero overlap. We do include the axismmetric component of the centrifugal distortion as described in Appendix \ref{math}. Equation \ref{TV} can be solved separately for each value of $m$, and modes have simple longitudinal eigenfunctions that are proportional to $e^{i m \phi}$. Including the Coriolis and the full centrifugal forces breaks this symmetry, so that all values of $m$ must be solved simultaneously. 

In our calculations, we only include the lowest value of $\ell_t$ that contributes to the coupling between two modes, as higher values of $\ell_t$ have smaller values of $U_{\ell_t}$. For example, $\ell=1$ modes are coupled with $\ell=3$ modes through the $\ell_t=2$ component of the tidal potential, while they are coupled with $\ell=6$ modes by the $\ell_t=5$ component of the tidal potential. We note that $\ell=0$ modes do not couple with other $\ell=0$ or $\ell=1$ modes, but they do couple with $\ell \geq 2$ modes through $\ell_t \geq 2$. Dipole modes ($\ell=1$) couple with each other through $\ell_t=2$ and with $\ell=2$ modes through $\ell_t =3$. The coupling with odd values of $\ell_t$ is very important because it breaks the symmetry between the ${\rm L}_1$ and ${\rm L}_3$ sides of the star, allowing modes to be tidally trapped on one side of the star.

Our approach has important advantages over some other methods of dealing with tidal or centrifugal distortion. Many papers only consider the frequency corrections due to the coupling of a mode with itself, which leads to rotational or tidal frequency splitting of mode multiplets. This is adequate if the off-diagonal terms in equation \ref{TV} (which represent coupling between modes of different $\ell$ or $n$) are much smaller than the differences between mode frequencies, but very often this is not the case. In particular, g~modes are often closely spaced in frequency, so they can be strongly coupled with other g~modes or a nearby f~mode or p~mode (i.e., mixed modes). Other works consider second-order perturbation theory that accounts for the lowest order mixing between different modes, such that nearly degenerate modes (i.e., modes in an avoided crossing) suffer the most mixing. While this approach is better, it is inadequate if networks of modes are strongly coupled, which our method can account for. Finally, we note that all of these methods can also be used to account for the effects of rotation or magnetic fields (e.g., \citealt{shibahashi1993,reese2010}). Coupling between modes of different $\ell$ also occurs in rotating stars, but the axisymmetry of that problem means that the mode amplitude is not modulated with rotational phase. 

\subsection{Application to Stellar Models}
\label{application}

In practice, we must truncate the set of basis modes used in our solutions. For the results presented below, we include modes with frequencies surrounding those of the observed pulsations, as it is most important to include modes in this frequency range to capture avoided crossings between modes of similar frequency. We include values of $\ell$ in the range $0 \leq \ell \leq 10$, finding no significant differences if higher values of $\ell$ are included, hence our results appear to be converged.

To compute our basis modes, we construct stellar models of each of the systems discussed in Section \ref{results} using the MESA stellar evolution code \citep{paxton:11,paxton:13,paxton:15,paxton:18,paxton:19}. The models have masses, radii, and temperatures within the measurement uncertainties for each system listed in Table \ref{table1}. We then compute non-adiabatic pulsation modes using the GYRE pulsation code \citep{townsend:13}. We limit the frequency range to within a factor of roughly 2 of the observed value of $f_{\rm max}$ for each system. Computations with slightly different frequency ranges do not significantly alter the results for the modes discussed below, so we consider our calculations to be converged and to have a sufficiently complete set of basis modes. 

Using the oscillation modes of the spherical stellar models as basis functions, we then compute the overlap integrals (see Appendix \ref{math}) that define the elements of the matrix equation \ref{sum}. In this step, we only include overlaps between the real components of each eigenfunction, which is a good approximation because the imaginary components are small in the stellar interior where most of the coupling occurs. We also use only the real part of the eigenfrequency, which is always much larger than the imaginary part. Finally, we solve the matrix equation \ref{TV} for the eigenfrequencies $\omega$ and expansion coefficients $a_\alpha$ of the modes of the distorted star. Using the expansion coefficients $a_\alpha$ and temperature perturbations $\Delta T_\alpha$ of each basis mode, we can compute each mode's flux perturbation across the surface of the star.

\begin{table}
	\centering
\begin{tabular}{c|c|c|c}
\quad & {\bf HD~74423} & {\bf CO~Cam} & {\bf TIC~63328020} \\ 
\hline
\hline
$P$ (d) & $1.580723$ & $1.27099$ & $1.105769$ \\
$M_1 ({\rm M}_\odot)$ & $2.1 \pm 0.1$ & $1.48^{+0.02}_{-0.01}$ & $2.5 \pm 0.2$ \\
$M_2 ({\rm M}_\odot)$ & $2.0 \pm 0.1$ & $0.86 \pm 0.02$ & $1.07 \pm 0.06$ \\
$R_1 ({\rm R}_\odot)$ & $3.3 \pm 0.1$ & $1.83 \pm 0.01$ & $3.1 \pm 0.1$ \\
$R_2 ({\rm R}_\odot)$ & $3.2 \pm 0.1$ & $0.84 \pm 0.02$ & $2.06 \pm 0.06$ \\
$T_1$ (K) & $7900 \pm 150$ & $7080 \pm 80$ & $8200 \pm 450$ \\
$T_2$ (K) & $7600 \pm 200$ & $5050 \pm 150$ & $5600 \pm 250$ \\
$i$ (deg) & $33 \pm 2$ & $48.9 \pm 1.0$ & $79.1 \pm 0.6$ \\
$R_1/R_L$ & $>0.95$ & $0.65 \pm 0.02$ & $>0.95$ \\
$R_1/a$ & $0.36 \pm 0.02$ & $0.28 \pm 0.01$ & $0.45 \pm 0.03$ \\
$f_{\rm max} \, ({\rm d}^{-1})$ & 8.76 & 13.38 & 21.10 \\
$n_{\rm max}$ & 2 & 1 & 5 \\
$\Delta R_{\rm tide}/\Delta R_{\rm cen}$ & $1.5 \pm 0.1$ & $1.10 \pm 0.02$ & $0.9 \pm 0.1$ \\
\end{tabular}
\caption{\label{table1} Properties of known tidally tilted pulsators, taken from \citealt{handler2020} and private communication from Kahraman Ali\c{c}avu\c{s}, \citealt{kurtz2020}, and Rappaport et al., in preparation. The rows are orbital period $P$, mass of pulsating star $M_1$, mass of companion star $M_2$, radius of pulsating star $R_1$, radius of companion star $R_2$, temperature of pulsating star $T_1$, temperature of companion star $T_2$, orbital inclination $i$, Roche filling factor of pulsating star $R_1/R_L$, primary radius divided by semi-major axis $R_1/a$, the frequency of the highest amplitude pulsation mode $f_{\rm max}$, and its most probable radial order $n_{\rm max}$ (assuming a radial mode).}
\end{table}

\subsection{Amplitude and Phase Modulation}
\label{amp}

The main purpose of our tidally coupled mode computations is to predict the modulation of the mode's amplitude and phase throughout the orbit.
To do this, we compute the Lagrangian flux perturbation for each mode, $\Delta F_\alpha$, and we assume the disc-integrated luminosity perturbation is due solely to flux perturbations, i.e., we neglect the surface area and surface normal perturbations. Our non-adiabatic computations include both a real and an imaginary component of the mode eigenfunction, hence the surface flux perturbation contains a real and imaginary part which affects the mode phase variation.


To compute the mode's observed amplitude and phase, we must compute its disc-integrated luminosity fluctuation as the observer's viewing angle changes throughout the orbit. Appendix \ref{nonaxisymmetric} describes the details of the calculation, here we provide the basic method and results.
We project the mode flux perturbation $\Delta F(\theta,\phi)$ onto spherical harmonics,
\begin{equation}
    \label{dfdecomp}
    \Delta F(\theta,\phi) = \sum_\ell \Delta F_\ell  Y_{\ell m}(\theta,\phi) \, .
\end{equation}
The flux perturbation for each value of $\ell$ is
\begin{equation}
\label{dfl}
    \Delta F_\ell = 2 \pi \int^{\pi}_{0} \Delta F(\theta,\phi)  \, Y_{\ell m}^*(\theta,\phi) \sin \theta d \theta \, .
\end{equation}
Here we have already integrated over $\phi$, and have assumed $\Delta F(\theta, \phi)$ is proportional to $e^{i m \phi}$, as it is for our calculations that are axisymmetric about the tidal axis.


The observed luminosity amplitude is found by a rotation to the observer's frame and then integrating over the disc of the star, as described in Appendix \ref{nonaxisymmetric}. The  observed luminosity fluctuation as a function of orbital phase $\phi_{\rm o}$ is 
\begin{align}
\label{dlobs}
\Delta L(\phi_{\rm o}) \propto &\sum_{\ell} \bigg(\sqrt{2 \ell+1} \, b_\ell \, \Delta F_\ell \nonumber \\ &\times \sum_{m_{\rm s}=-\ell}^{\ell} d^\ell_{m_{\rm s},0}(-\pi/2) d^\ell_{0,m_{\rm s}}(i_{\rm o}) e^{-im_{\rm s} \phi_{\rm o}} \bigg) \, .
\end{align}
Here, $\phi_{\rm o} = \Omega t$ is the orbital phase, and the Wigner coefficients $d^\ell_{m,m'}$ are used to convert the coordinate $(\theta,\phi)$ to spherical coordinates associated with the orbital axis, and then to the axis associated with the line of sight, which is inclined to the orbital axis by inclination angle $i_{\rm o}$. Equation \ref{dlobs} is similar to equation 3 of \cite{handler2020}, though note that the angle between the tidal axis and the orbital axis is $\beta \! = \! -\pi/2$, assuming the inclination angle $i_{\rm o}$ is defined to be positive. The factor $b_l$ is obtained by integrating the flux perturbation over the disc of the star,
\beq
\label{bl}
b_\ell = \int^{1}_0 P_{\ell}(\mu) h(\mu) \mu d \mu \, .
\eeq
where $\mu = \cos \theta$ and $h$ is a limb-darkening function. We adopt a linear limb-darkening model with $u=0.4$ (see \citealt{townsend:03}), appropriate for stars of $T\approx 7500 \, {\rm K}$ for the \textit{TESS} bandpass \citep{claret2017}. Different limb darkening laws or coefficients have only a small effect on our results. 

Recall that $\Delta F$ is generally complex, as are the coefficients $\Delta F_\ell$ of its decomposition. The amplitude of the pulsation as a function of orbital phase is then 
\beq
A_{\rm mode}(\phi_{\rm o}) = | \Delta L(\phi_{\rm o})| \, ,
\eeq
while the phase of the pulsation is 
\beq
\phi_{\rm mode}(\phi_{\rm o}) = {\rm atan2} \bigg( \frac{ -{\rm Im} \big[ \Delta L(\phi_{\rm o}) \big] }{ {\rm Re} \big[ \Delta L(\phi_{\rm o}) \big]} \bigg) \, .
\eeq

\section{Results for Tidally Tilted Pulsators}
\label{results}


In the discovery paper for HD~74423, \citet{handler2020} described the star as a ``single-sided pulsator'', given that they were able to show that the pulsation was largely trapped on either the L$_1$ or L$_3$ side of the primary star. \citet{kurtz2020} continued with that nomenclature for CO~Cam, which they were able to show has at least four pulsation modes trapped largely on the L$_1$ side of the primary star. However, TIC~63328020 (Rappaport et al., in preparation) has a single $\ell =1, m= -1$ mode that is not strongly trapped and is not technically ``single-sided''. We therefore introduce here a preferable descriptive term for this new class of stars, naming them ``tidally tilted pulsators'', of which single-sided pulsators are a sub-class. The tidally tilted pulsator name applies generally to stars with pulsations whose axis has been tilted away from the rotation axis due to tidal distortion, similar to the oblique pulsators whose pulsation axis has been tilted by magnetic fields.

Thus, we apply here our calculations to the three currently known tidally tilted pulsators discussed in the introduction: HD~74423, CO~Cam, and TIC~63328020. Table \ref{table1} lists the properties of each system. They are all $\delta$~Sct  pulsators in binaries with periods under two days, though notably HD~74423 and TIC~63328020 are nearly Roche lobe filling while CO~Cam is not. Each system pulsates in a small number of detected modes which show large amplitude and phase modulation over the course of the orbit, which is the clear signature that the symmetry axis of these modes is tilted away from the pulsator's spin axis, i.e., they are tidally tilted pulsators. The pulsation modes are all low-order acoustic or fundamental modes, as indicated by the approximate radial order $n_{\rm max}$ of the nearest dipole mode to the highest amplitude pulsation in each system. In fact, the pulsations in HD~74423 and CO~Cam are mixed modes because they have acoustic character in the envelope, but gravity mode character in the core.

\subsection{HD~74423}

\begin{figure}
\includegraphics[scale=0.36]{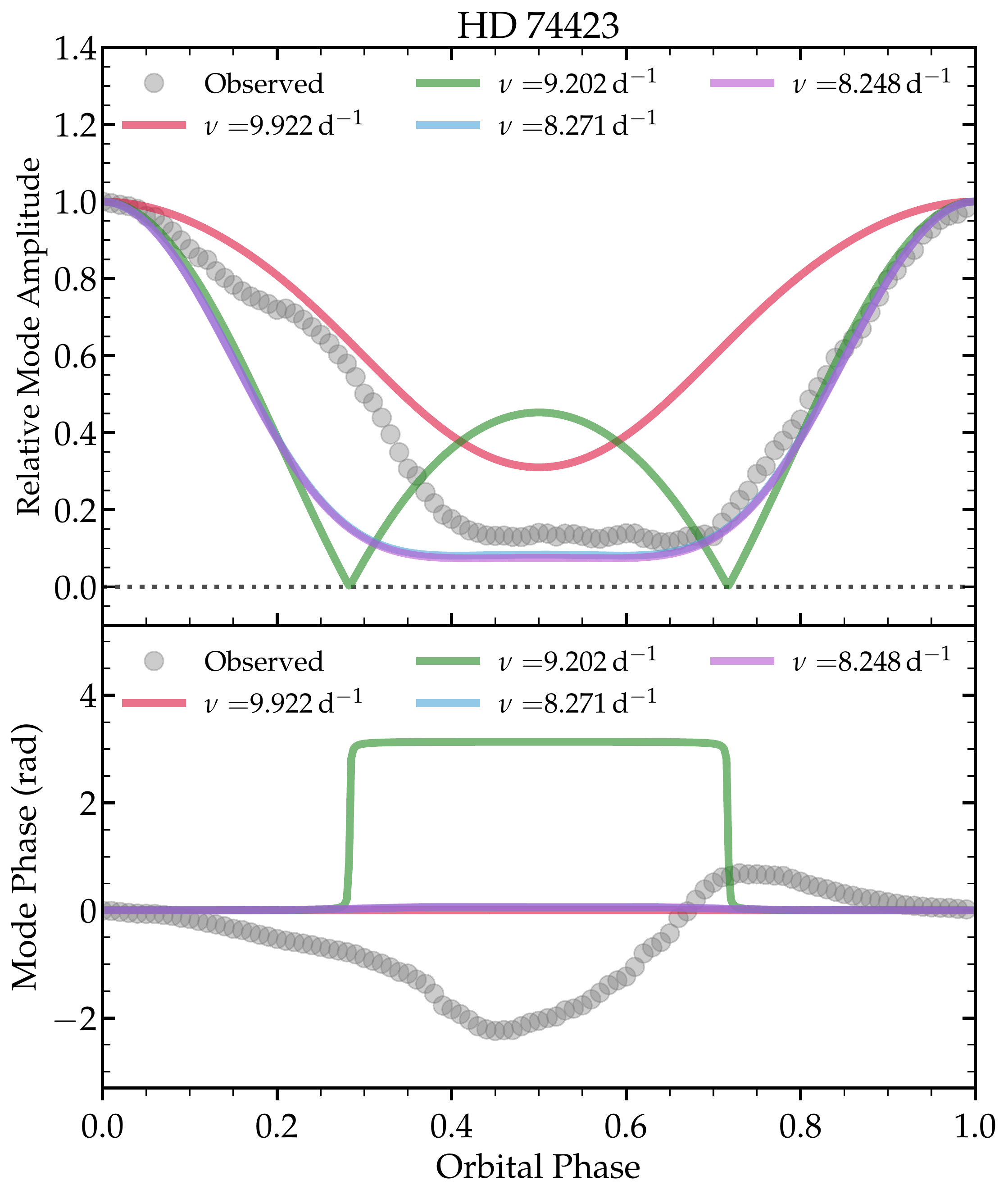}
\caption{\label{TIC35} 
Top: Amplitude variations of axisymmetric pulsation modes of HD~74423, as a function of orbital phase. Gray circles are the observed amplitudes and phases from \citealt{handler2020} after applying some smoothing, while each colored line is the prediction from one of the modes of our model, with frequencies indicated in the legend. The corresponding surface flux perturbations as a function of tidal latitude are shown in Fig. \ref{TIC35flux}. These calculations assume an orbital inclination of $i=33$ deg relative to the line of sight. Bottom: Corresponding phase variations of the pulsation modes over the orbital cycle.
}
\end{figure}

We begin with an analysis of HD~74423, which exhibits a single pulsation mode with a frequency of $8.76 \, {\rm d}^{-1}$. The mode amplitude and phase are modulated strongly over the orbit, as shown in Fig. \ref{TIC35}. The orbital phase is defined as zero at the photometric minimum of the ellipsoidal distortion, which likely occurs when the L$_1$ side of the pulsating star is closest to the line of sight. We can see that the pulsation amplitude has a maximum at this phase, and a deep flat-bottomed minimum near orbital phase 0.5. 

\begin{figure}
\includegraphics[scale=0.36]{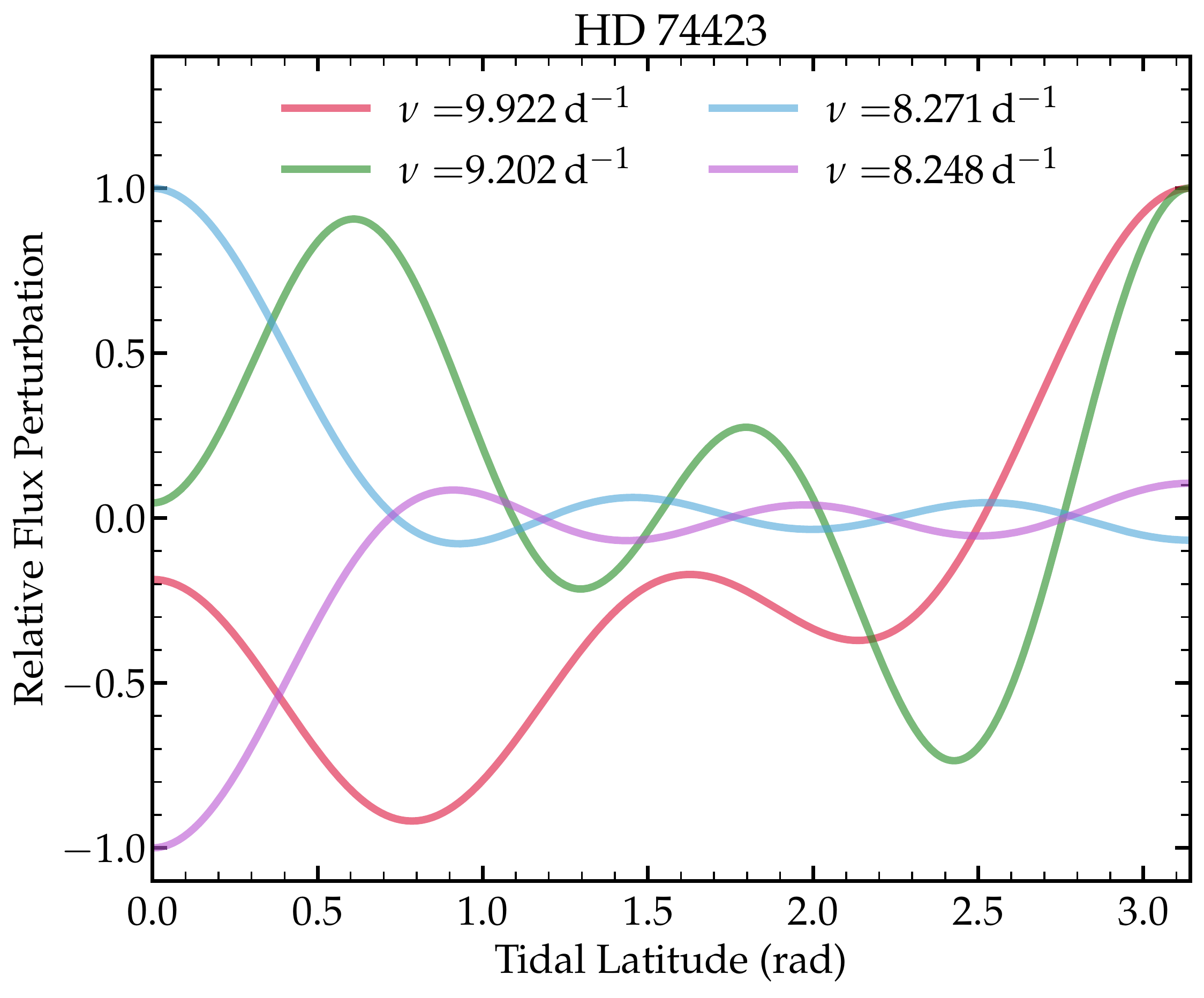}
\caption{\label{TIC35flux} 
Real component of the surface flux perturbations as a function of tidal latitude for the four modes shown in Fig. \ref{TIC35}. The modes with $\nu=8.271 \, {\rm d}^{-1}$ and $\nu=8.248 \, {\rm d}^{-1}$ have flux perturbations that peak on the L$_1$ side of the star, causing the observed luminosity fluctuations to be largest at orbital phase zero, when the L$_1$ side of the star is visible.
}
\end{figure}

The colored lines in Fig. \ref{TIC35} show the predicted amplitude and phase modulation of four modes with frequencies comparable to that observed. To pick which modes to plot, we selected only $m=0$ modes with frequencies in the range $7 \, {\rm d}^{-1} \leq f \leq 10 \, {\rm d}^{-1}$, whose amplitude at orbital phase 0.5 is less than 50\% the amplitude at orbital phase 0, and we only plot the four modes with the highest luminosity perturbation. Because our calculation includes basis modes up to $\ell=10$, there is a dense spectrum of high-$\ell$ g~modes in this frequency range that are largely trapped in the core. These g~modes would be difficult to detect because of their high values of $\ell$ and high mode inertias. Our selection favors the modes most likely to be observed, which encouragingly are within a few percent of the observed mode frequency of HD~74423.

\begin{figure}
\centering
\includegraphics[width=0.48\columnwidth]{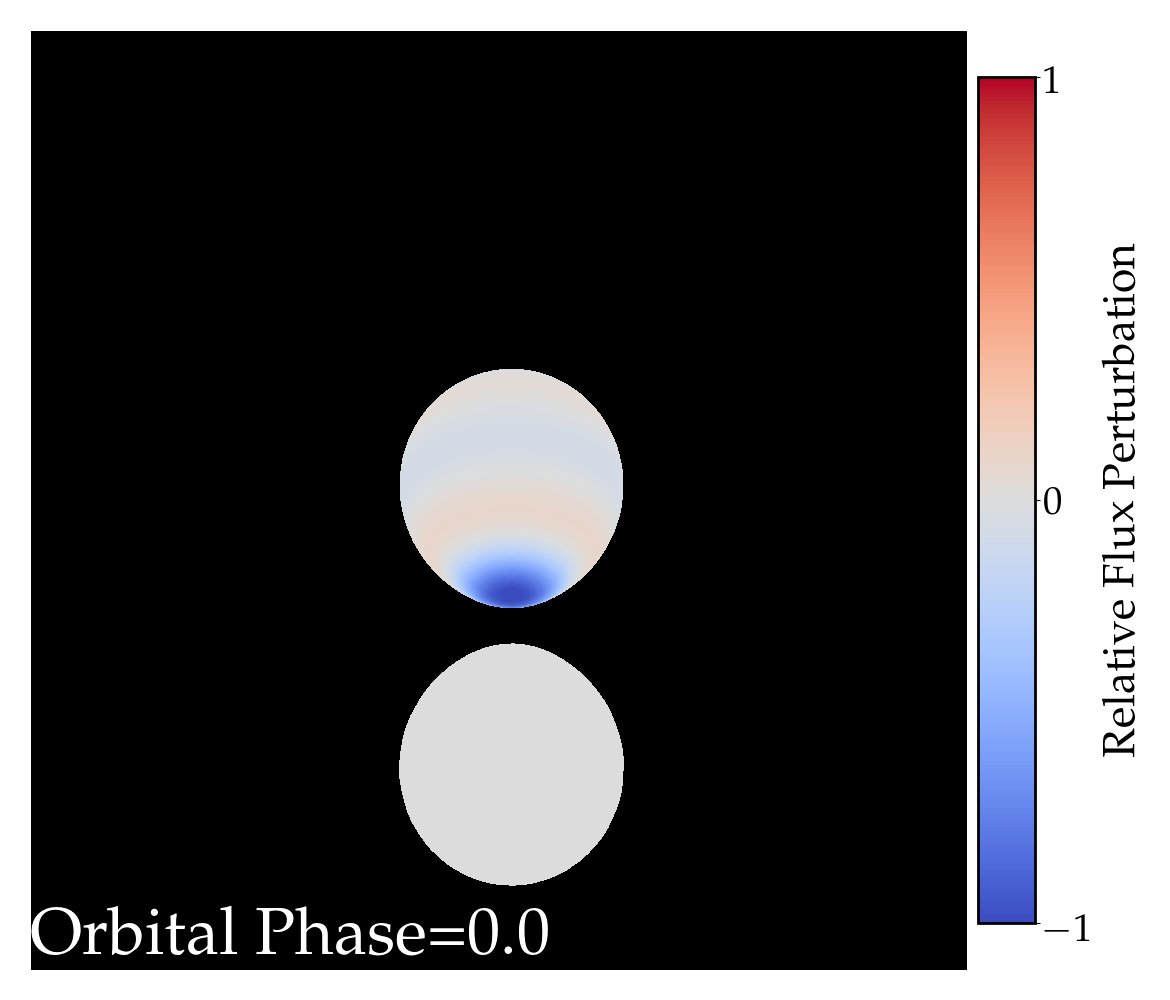}
\includegraphics[width=0.48\columnwidth]{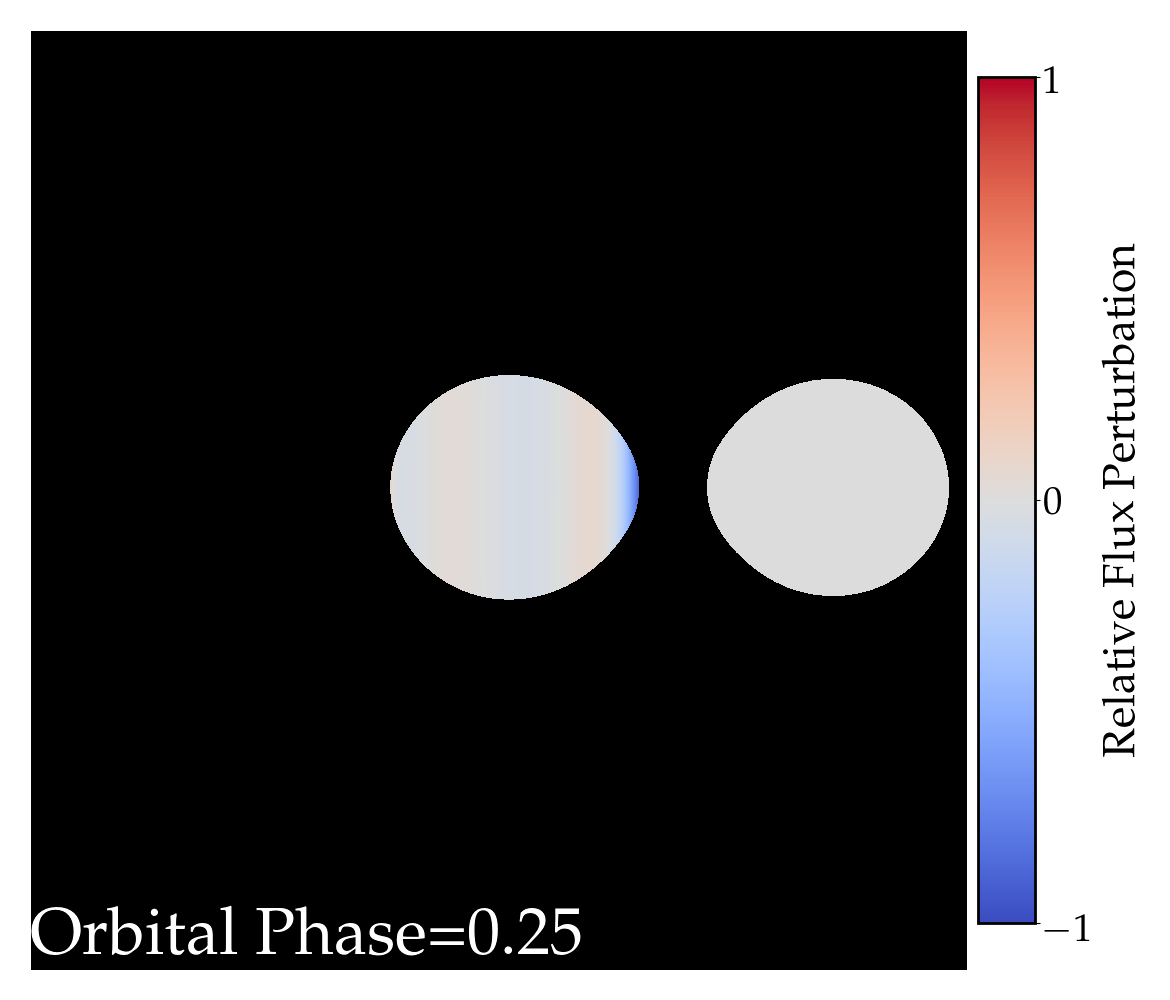}
\includegraphics[width=0.48\columnwidth]{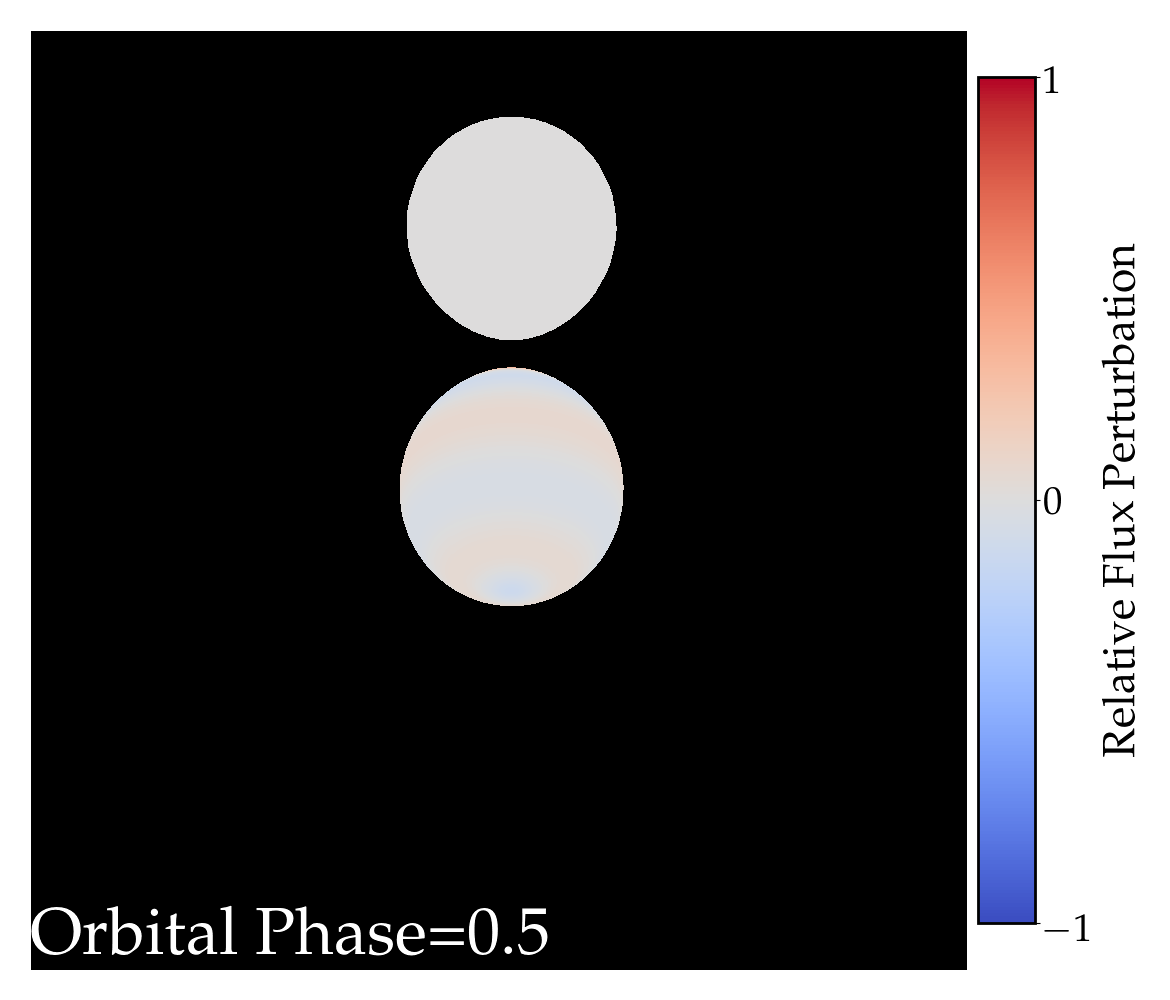}
\includegraphics[width=0.48\columnwidth]{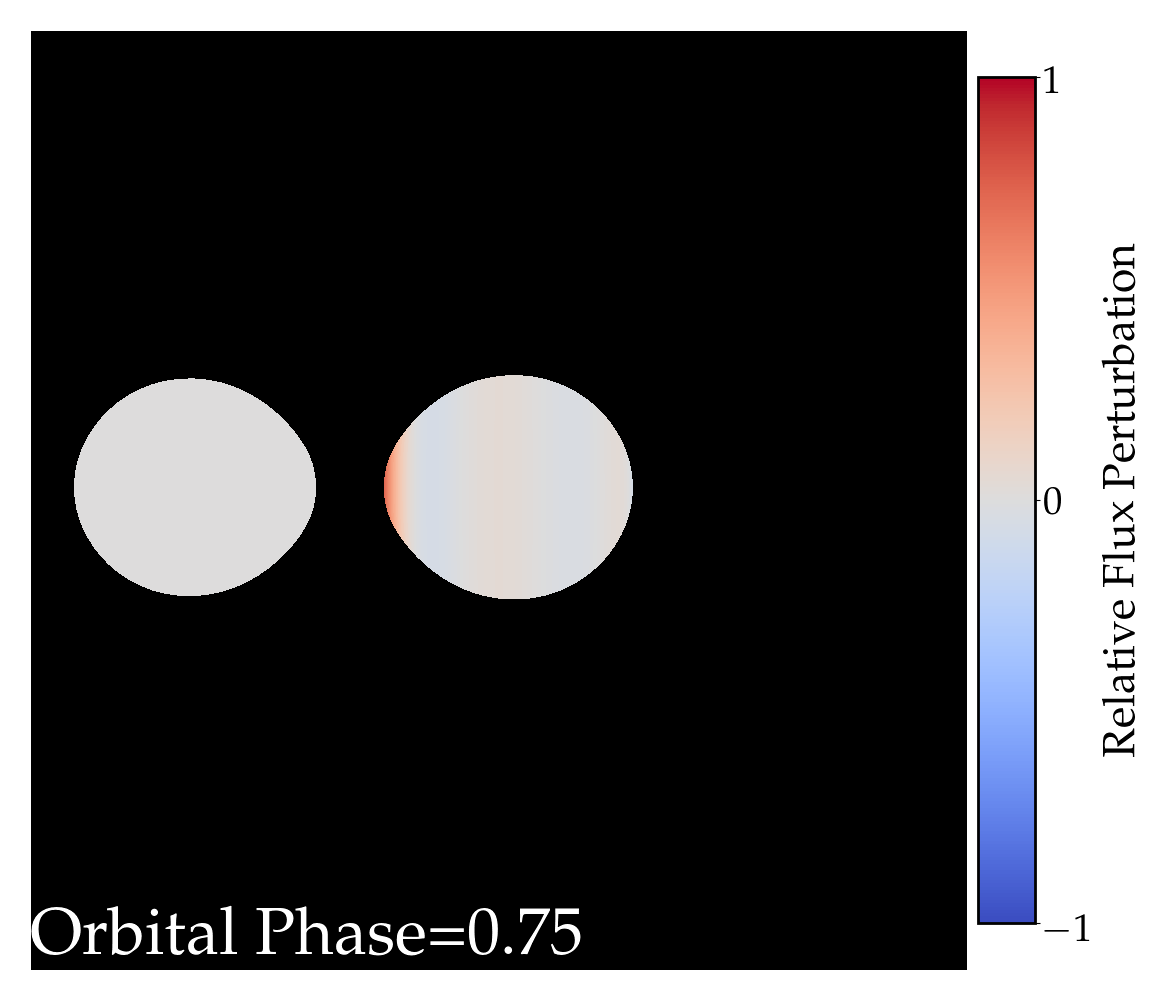}
\caption{\label{fig:TIC35fig}
A three dimensional model of the pulsation at $\nu = 8.248 \, {\rm d}^{-1}$ in HD~74423. The system is viewed from an inclination of $i=33$ degrees, with parameters from Table \ref{table1}, and is fixed on the center of mass of the pulsator. See the electronic version of this article for an animated version of this figure.
}
\end{figure} 

As can be seen, the predicted orbital amplitude modulation of these four modes resembles the observed mode in HD~74423, with a very good match for the two modes at $\approx \! 8.25\,{\rm d}^{-1}$. The corresponding flux perturbation as a function of tidal latitude for each mode is shown in Fig. \ref{TIC35flux}.
The two modes at $\approx \! 8.25\,{\rm d}^{-1}$ have nearly identical flux distributions (apart from a different sign) due to the fact that they are hybrid modes in an avoided crossing as discussed below. Their flux perturbations peak on the L$_1$ side and are small at mid-latitudes, i.e., they are trapped at the tidal pole. For this reason, their observed luminosity amplitude peaks at orbital phase zero, when the L$_1$ side of the star is visible. The other modes above $9 \, {\rm d}^{-1}$ actually have flux perturbations that peak on the L$_3$ side of the star, but cancellation between positive and negative flux perturbations causes a lower luminosity amplitude near orbital phase 0.5.

Fig. \ref{fig:TIC35fig} shows a three-dimensional model of the pulsation at $\nu=8.248 \, {\rm d}^{-1}$ in HD~74423, illustrating the larger observable pulsation amplitude at an orbital phase of zero, due to the tidal trapping of the mode on the L$_1$ side of the star.

There is a noticeable asymmetry in the observed amplitude modulation that cannot be reproduced with our axisymmetric modes. We attribute this asymmetry to the Coriolis force (not included in our models), which we discuss further in Section \ref{disc}. Otherwise, however, the shape and minimum of the theoretically predicted amplitude modulation for the two modes at $\approx \! 8.25\,{\rm d}^{-1}$ matches the observations remarkably well. We cannot reliably predict which of these modes will grow to the largest amplitude and actually be observed. The mode at $8.248 \, {\rm d}^{-1}$ has the largest normalized surface flux perturbation and has the largest fraction of its energy in the acoustic cavity, so it is probably the best candidate.

The bottom panel of Fig. \ref{TIC35} shows the observed and predicted mode phases as a function of orbital phase. Here the match is not as satisfactory. The models predict a phase shift of $\simeq \! 0$ or $\simeq \! \pi$ over half the orbit, depending on the exact shape of the flux perturbation in Fig. \ref{TIC35flux}. Note that a phase shift of $\approx + \pi$ is almost identical to a phase shift of $\approx - \pi$, which is similar to what is observed. The predicted phase shift can also depend on the orbital inclination, which is not yet precisely determined for HD~74423. We also note that the observed phases are obtained by fitting wavelets to a small fraction of the orbital cycle, so they are smoothed relative to the actual amplitude and phase variations. Applying some smoothing to modes with phase shifts of $\simeq \! \pi$, like the mode with $\nu = 9.2 \, {\rm d}^{-1}$, would better match the data. 


A few modes with similar eigenfunctions and closely spaced frequencies, like the two modes with $\nu \approx \! 8.25\,{\rm d}^{-1}$ shown in Fig. \ref{TIC35}, are common in our models. These mode clusters are a clear signature of mixing between modes of different $\ell$ due to tidal distortion. Normally, axisymmetric modes of the same $\ell$ (i.e., similar surface flux distributions) are spaced by the large frequency spacing, which in HD~74423 is $\Delta \nu \simeq 2.5 \, {\rm d}^{-1}$. The low radial order $(n \! \sim \! 1)$ of the mode in HD~74423 implies an even larger frequency spacing than the asymptotic spacing $\Delta \nu$.  Coupling between acoustic modes may alter frequency spacings but does not increase the p~mode density and does not typically result in clusters of modes with finely spaced frequencies.

Instead, the origin of this phenomenon is mode mixing between p~modes and g~modes. The g~modes whose frequencies are similar to p~modes undergo avoided crossings with the p~modes, producing the strongest mode mixing, resulting in a cluster of hybrid modes around each p~mode. This phenomenon has also been observed in Saturn's pulsation modes \citep{fullersatb:14} due to rotational mode mixing and can occur in mixed modes in red giant stars \citep{deheuvels:17}. In HD~74423, evidently only one mode is excited, but a cluster of hybrid modes is observed in CO~Cam as discussed below.

\subsection{CO~Cam}

\begin{figure}
\includegraphics[scale=0.36]{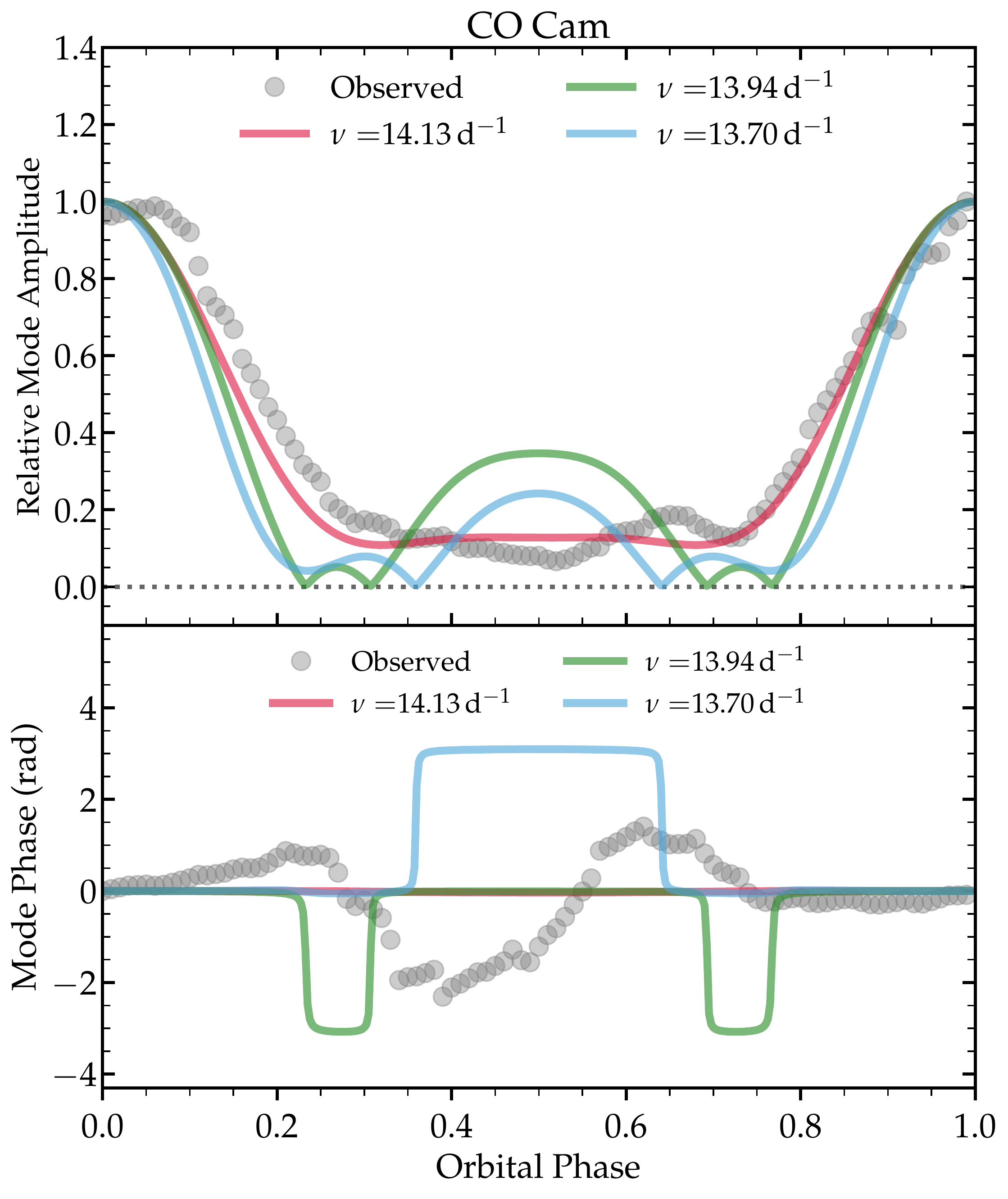}
\caption{\label{COCam} 
Top: Same as Fig. \ref{TIC35}, but for the modes of CO~Cam. We show only the mode $\nu_1$ \citep{kurtz2020}, but its three other modes have similar frequencies and amplitude modulation. The predictions assume an orbital inclination of $i = 49$ deg. The corresponding flux perturbations across the surface of the star are shown in Fig. \ref{COCamflux}. Bottom: Mode phase variations as a function of orbital phase. 
}
\end{figure}

In many ways, CO~Cam exhibits similar behavior to HD~74423, as shown by the observed amplitude and phase modulations \citep{kurtz2020} in Fig. \ref{COCam} for its highest amplitude mode. Both the amplitude and phase modulation of this mode are remarkably similar to that in HD~74423, suggesting the same underlying cause. Unlike HD~74423, however, CO~Cam significantly underfills its Roche lobe, so it is clear that tidal trapping is not limited to the most tidally distorted stars. Another important difference is that CO~Cam exhibits \textit{four} modes with strong amplitude and phase modulation, with frequencies of $\nu_1 = 13.38 \, {\rm d}^{-1}$, $\nu_2 = 13.09 \, {\rm d}^{-1}$, $\nu_3 = 13.78 \, {\rm d}^{-1}$, $\nu_4 = 14.11 \, {\rm d}^{-1}$. The amplitude modulation of each mode is similar, so we only plot observations for $\nu_1$ for simplicity. The observed phase variations, however, do show some differences between the modes. 

\begin{figure}
\includegraphics[scale=0.36]{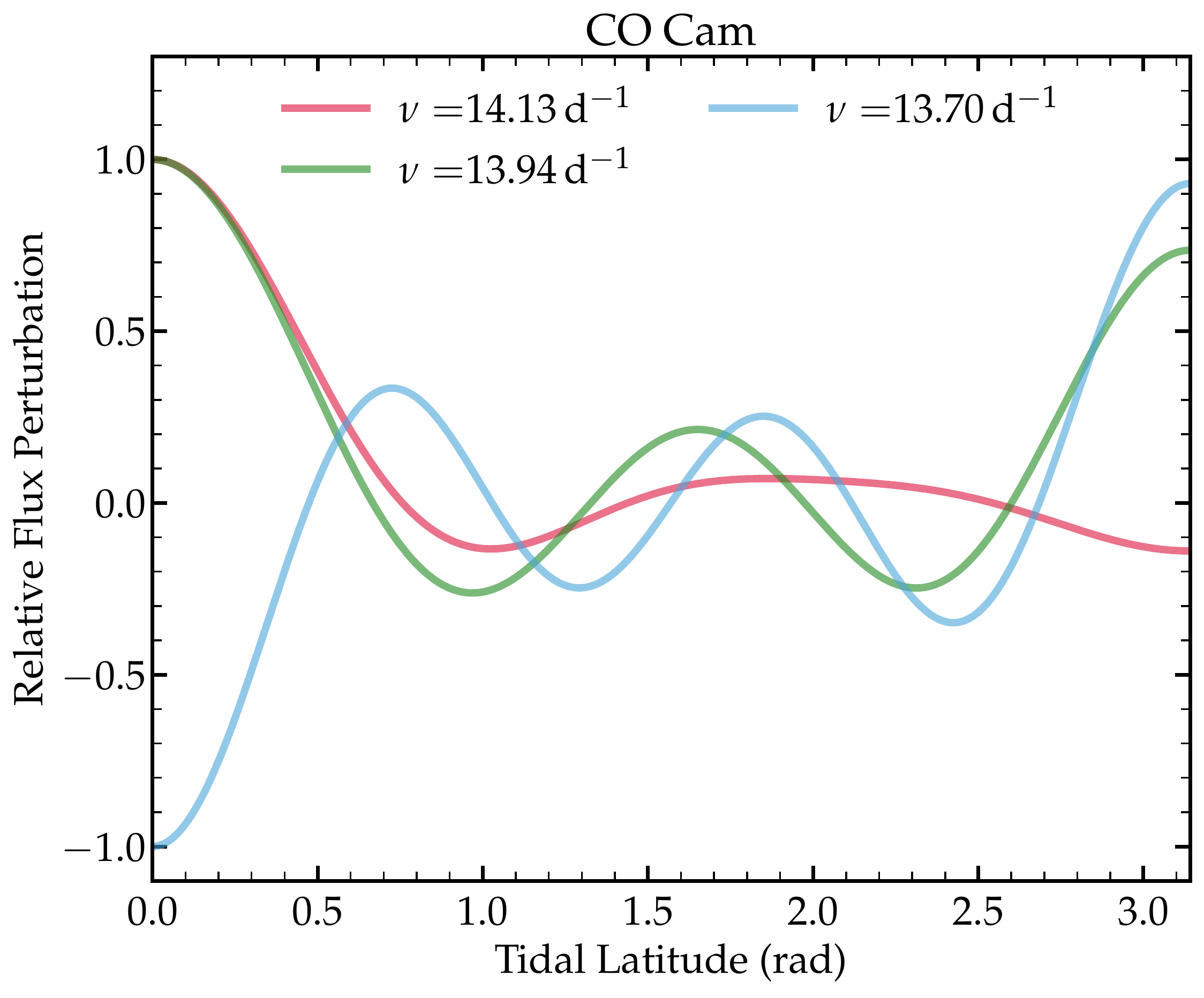}
\caption{\label{COCamflux} 
Real component of the relative flux perturbation as a function of tidal latitude for the modes of CO~Cam shown in Fig. \ref{COCam}. Though the fluxes peak on the L$_1$ side, significant power at mid and high latitudes allows for significant mode amplitudes at orbital phase 0.5. 
}
\end{figure}

The colored lines in Fig. \ref{COCam} are the predicted amplitude and phase modulations for modes selected in the same way as described for HD~74423, but this time in a frequency range  $11.5 \, {\rm d}^{-1} \leq f \leq 15.5 \, {\rm d}^{-1}$. Similar to HD~74423, the models for CO~Cam predict a few modes in this frequency range, closely spaced in frequency, each partially tidally trapped on the L$_1$ side of the star. The corresponding flux perturbations across the surface of the star are shown in Fig. \ref{COCamflux}. The amplitude of each mode peaks near orbital phase zero and is much smaller near orbital phase 0.5, broadly consistent with the observations. The match for the mode at $\nu = 14.13 \, {\rm d}^{-1}$ is especially good. However, for two of the predicted modes, the amplitudes have a small peak at orbital phase 0.5, in contrast to the flat-bottomed minima that are observed. We note that the observed amplitudes near orbital phase 0.5 have significant uncertainty due to lower signal to noise, and they are also smoothed in time by the measurement technique. Accounting for smoothing that is intrinsic to the measurement process would produce a better fit, but may not totally resolve the differences. 

\begin{figure}
\centering
\includegraphics[width=0.48\columnwidth]{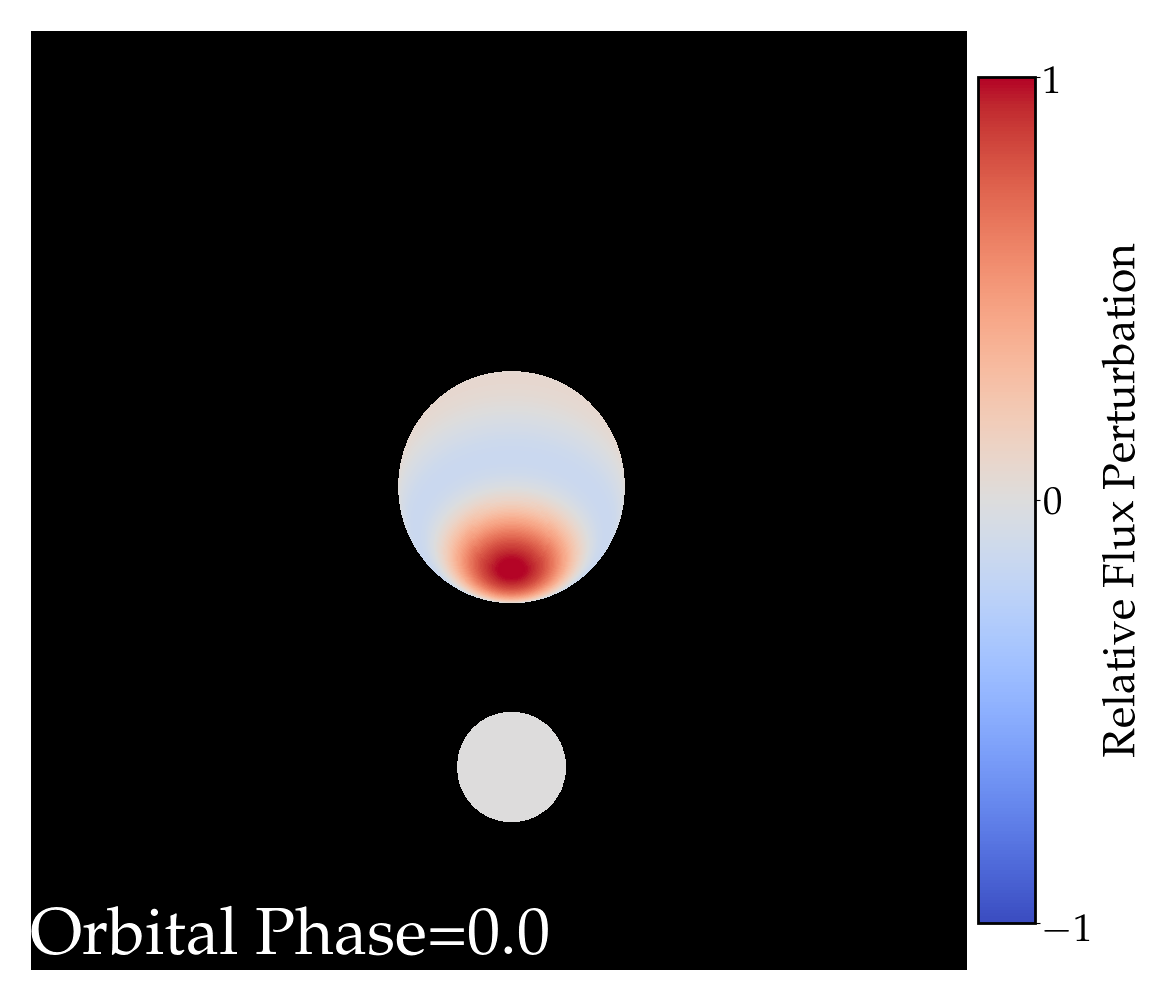}
\includegraphics[width=0.48\columnwidth]{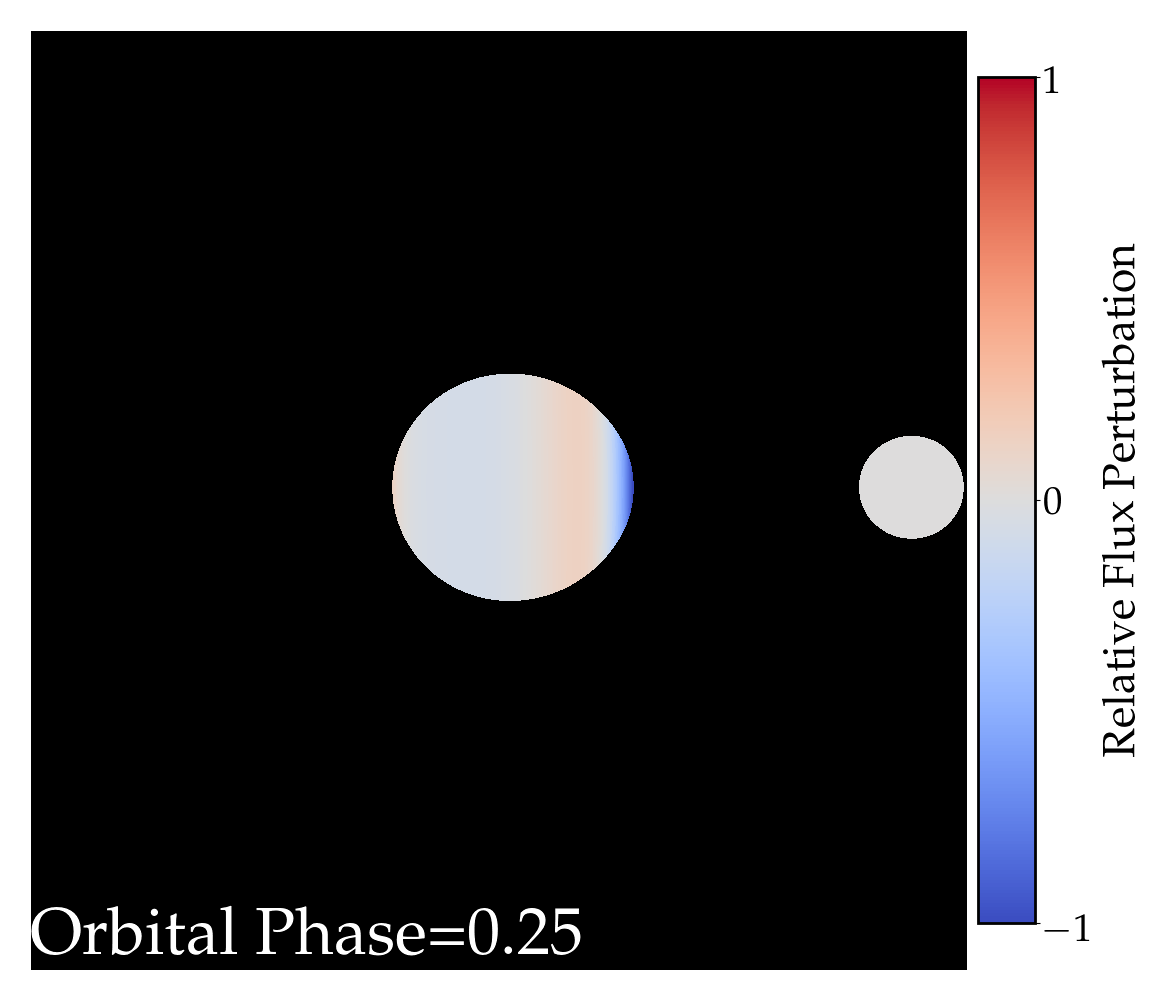}
\includegraphics[width=0.48\columnwidth]{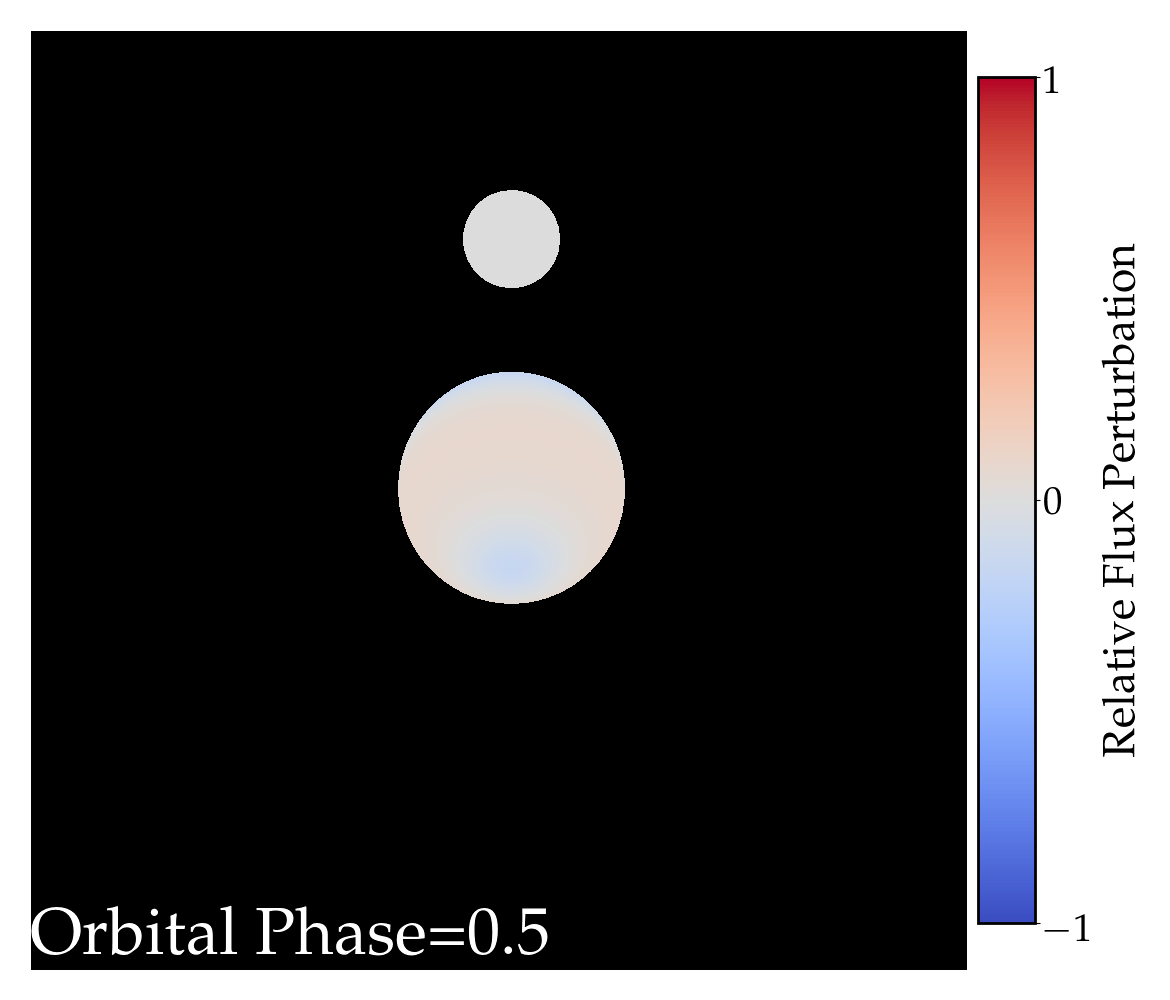}
\includegraphics[width=0.48\columnwidth]{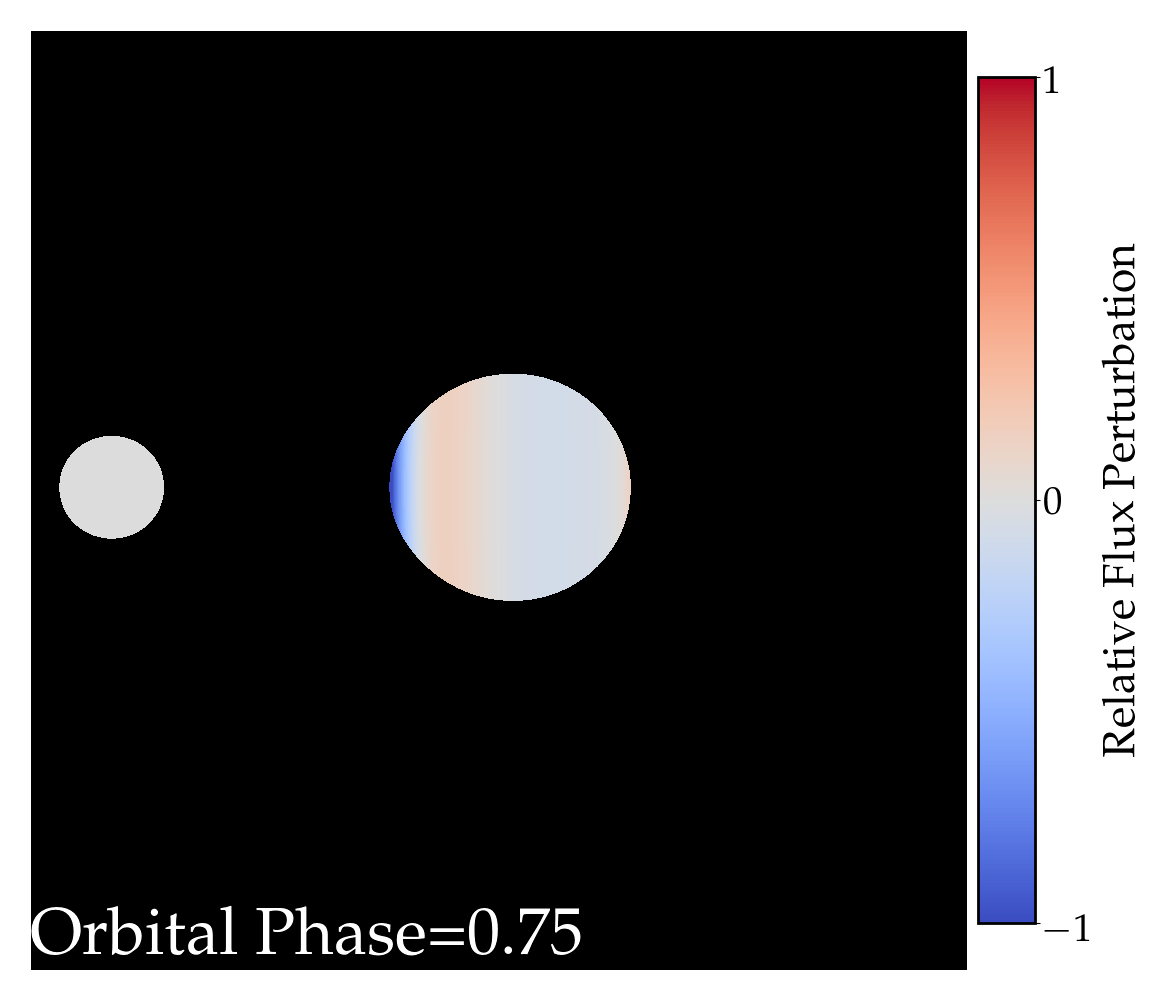}
\caption{\label{fig:COCamfig}
A three dimensional model of the pulsation at $\nu = 14.13 \, {\rm d}^{-1}$ in CO~Cam. The system is viewed from an inclination of $i=49$ degrees, with parameters from Table \ref{table1}, and is fixed on the center of mass of the pulsator. See the electronic version of this article for an animated version of this figure.
}
\end{figure}

Fig. \ref{fig:COCamfig} shows a three-dimensional model of the pulsation at $\nu=14.13 \, {\rm d}^{-1}$ in CO~Cam, which clearly illustrates the larger observable pulsation amplitude at an orbital phase of zero.

Another small difference between the models and the data are the mode frequencies. For this model, the predicted frequencies are a few percent larger than the observed frequencies. A slightly larger stellar radius than that of the model would produce better agreement. Alternatively, our neglect of centrifugal forces may be to blame, as they will systematically shift the non-radial p~modes to slightly smaller frequencies.

At first glance, the predicted phase modulations in Fig. \ref{COCam} appear somewhat discrepant from the observations. However, we again note that the models typically predict one or more phase shifts of $\pi$\,rad between orbital phase 0 and 0.5, similar to the observed phase shift.
Phase shifts of $\pi$\,rad usually occur near amplitude minima when the real part of the disc-integrated luminosity fluctuation passes through zero. Measuring such phase shifts for low-amplitude modes is extremely challenging because of the low signal and rapid variation of the phase. Indeed, the data in \cite{kurtz2020} does show large scatter when the modes have small amplitudes. We suspect that the gradual observed phase variations of less than $\pi$\,rad in Fig. \ref{COCam} are smoothed versions of the actual phase variations, which can contain multiple (and more sudden) jumps in phase by $\approx \pi$\,rad. We hope that more sensitive future measurements will be able to better constrain the actual mode phases for better comparison with the models. 

When constructing the models for HD~77423 and CO~Cam, we found that fairly small changes in the model (e.g., the star's radius) or the mass of the perturber (and therefore the strength of the tidal perturbation) could produce significant changes in the exact frequencies and amplitude variability of the predicted modes. We suspect this sensitivity arises from the avoided crossings between p~modes and g~modes that gives rise to hybrid modes, which is the source of the observed cluster of finely spaced modes with frequencies near 14 ${\rm d}^{-1}$. Small differences in the model can change which avoided crossings occur (i.e., the $\ell$ values for basis modes involved in the avoided crossings) which therefore changes the shape of the amplitude modulation over the orbital cycle. This behavior is most noticeable for CO~Cam because the envelope mode is a low-frequency fundamental mode, where the surrounding spectrum of g~modes is more dense than it is for higher frequency p~modes. This also helps explain why tidal trapping is so effective in CO~Cam despite its low Roche filling factor: the close frequency spacing allows for avoided crossings and strong coupling between modes despite the smaller tidal perturbation.

Nonetheless, we find that the models generally produce at least three modes near the observed frequency range whose amplitude is modulated in a manner qualitatively similar to the modes shown in Fig. \ref{COCam}. At least one mode with a flat-bottomed amplitude modulation and large surface flux perturbation (i.e., $\nu = 8.25 \, {\rm d}^{-1}$ for HD~74423 and $\nu = 14.13 \, {\rm d}^{-1}$ for CO~Cam) are robust features of the models. It is primarily the surrounding hybrid modes resulting from avoided crossings that are most sensitive to the details of the model. We therefore believe that the models are reproducing the correct basic behavior to explain the tidally trapped pulsations in CO~Cam.

\subsection{TIC~63328020}

\begin{figure}
\includegraphics[scale=0.36]{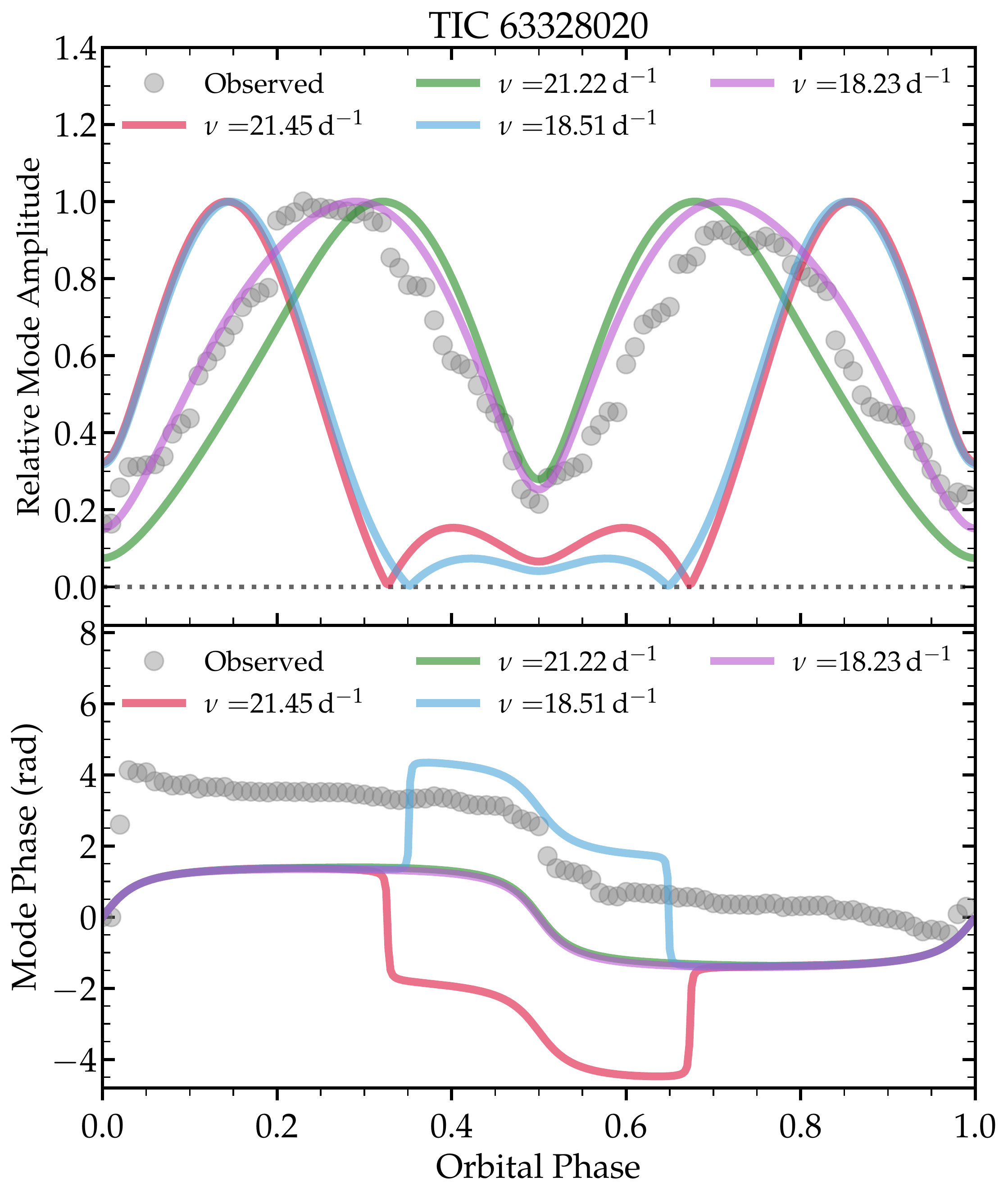}
\caption{\label{TIC63} 
Same as Fig. \ref{COCam}, but for the for modes of TIC~63328020 shown in Fig. \ref{TIC63flux}. The predictions assume an orbital inclination of $i = 79$ deg. Here the match is very good for two of the modes, including the match for the mode phase, provided a small shift in zero point of the mode's phase (see text). 
}
\end{figure}

The observed luminosity and phase variations of the third tidally tilted pulsator, TIC~63328020, are shown in Fig. \ref{TIC63}. In this system, the phenomenology is much different: the amplitude modulation of its primary mode at $\nu=21.1\, {\rm d}^{-1}$ exhibits a double-peaked structure, with peaks at orbital phases near 0.25 and 0.75, rather than a single peak at orbital phase 0. The amplitude modulation is approximately sinusoidal over the orbit. As in the previous two systems, the phase shifts by $\simeq \! \pi$\,rad over half the orbital phase, but with two sharp jumps at orbital phase 0 and 0.5.

\begin{figure}
\includegraphics[scale=0.36]{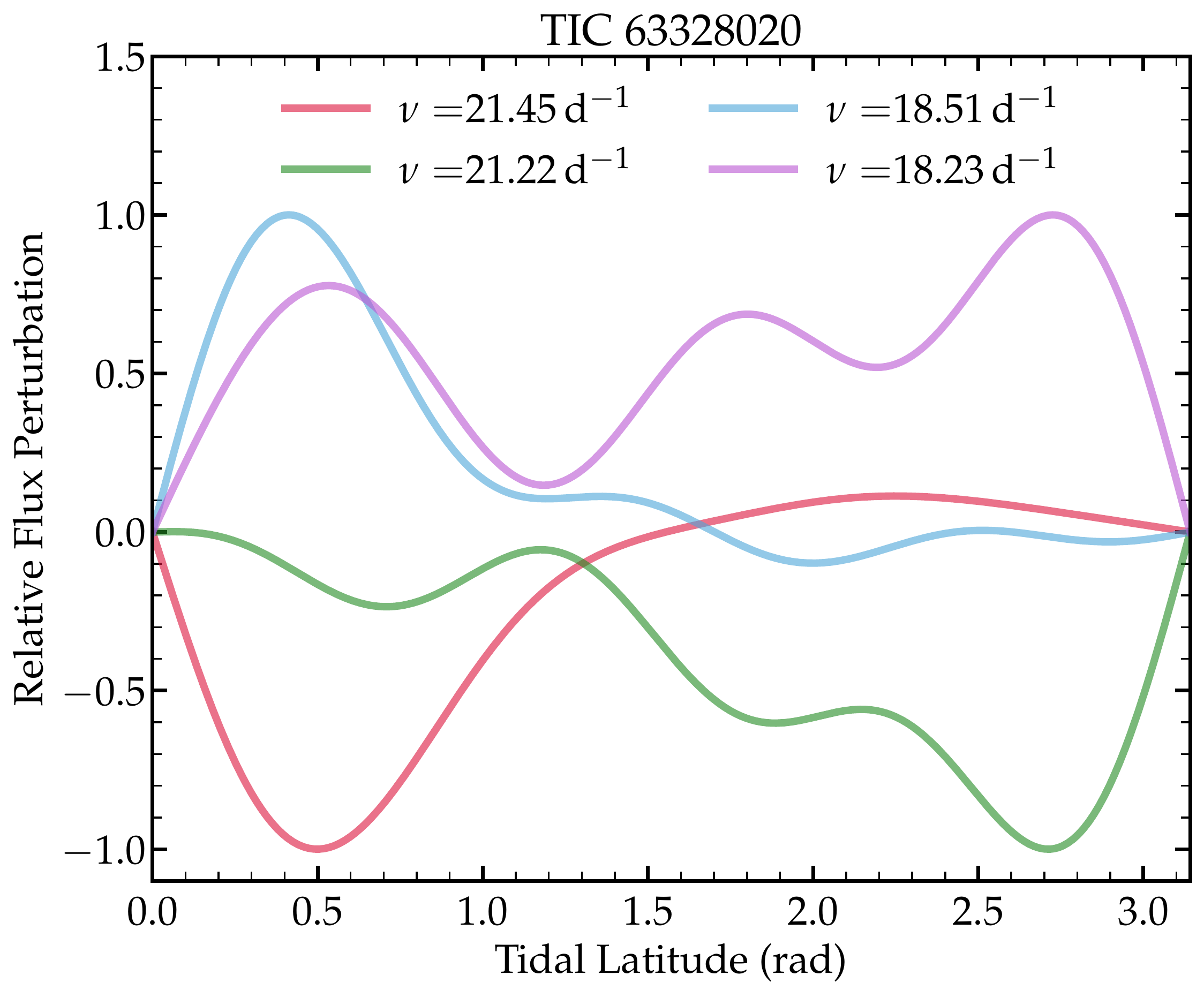}
\caption{\label{TIC63flux} 
Real component of the relative flux perturbation as a function of tidal latitude for the modes of TIC~63328020 shown in Fig. \ref{TIC63}. In this case, the modes have $m=-1$ longitudinal dependence.
}
\end{figure}

The different phenomenology of TIC~63328020 can be easily understood as the signature of non-axisymmetric $|m|=1$ tidally aligned modes. The four lines in Fig. \ref{TIC63} show the predicted amplitude and phase modulations for $|m|=1$ modes, two of which nicely track the observations. The corresponding flux perturbations for each mode are shown in Fig. \ref{TIC63flux}. In this case, we selected modes whose amplitude at orbital phase 0.5 is less than half the amplitude at orbital phase 0.25, and with frequencies in the range $18 \, {\rm d}^{-1} \leq f \leq 24 \, {\rm d}^{-1}$.

The predicted mode at $18.23 \, {\rm d}^{-1}$ best matches the observed amplitude modulation, while the predicted mode at $21.22 \, {\rm d}^{-1}$ is closer in frequency to the observed mode but not quite as good of a match in its amplitude modulation. Still, the overall agreement is impressive given that there are no adjustable parameters in the models. These results are less sensitive to the exact parameters of the model, likely because these higher frequency p~modes (radial order $n \approx 5$) do not undergo avoided crossings with g~modes.

In this case, the double-peaked structure is primarily the result of the changing viewing angle of a tidally aligned mode, rather than tidal trapping. At orbital phases 0 and 0.5, the mode amplitude is small because we are looking at tidal pole, and we see both the positive and negative flux perturbations produced at opposite tidal longitudes by the tidally aligned $|m|=1$ mode. At orbital phases 0.25 and 0.75, the amplitude is maximized because we only see positive/negative flux perturbations with little flux cancellation. 

\begin{figure}
\centering
\includegraphics[width=0.48\columnwidth]{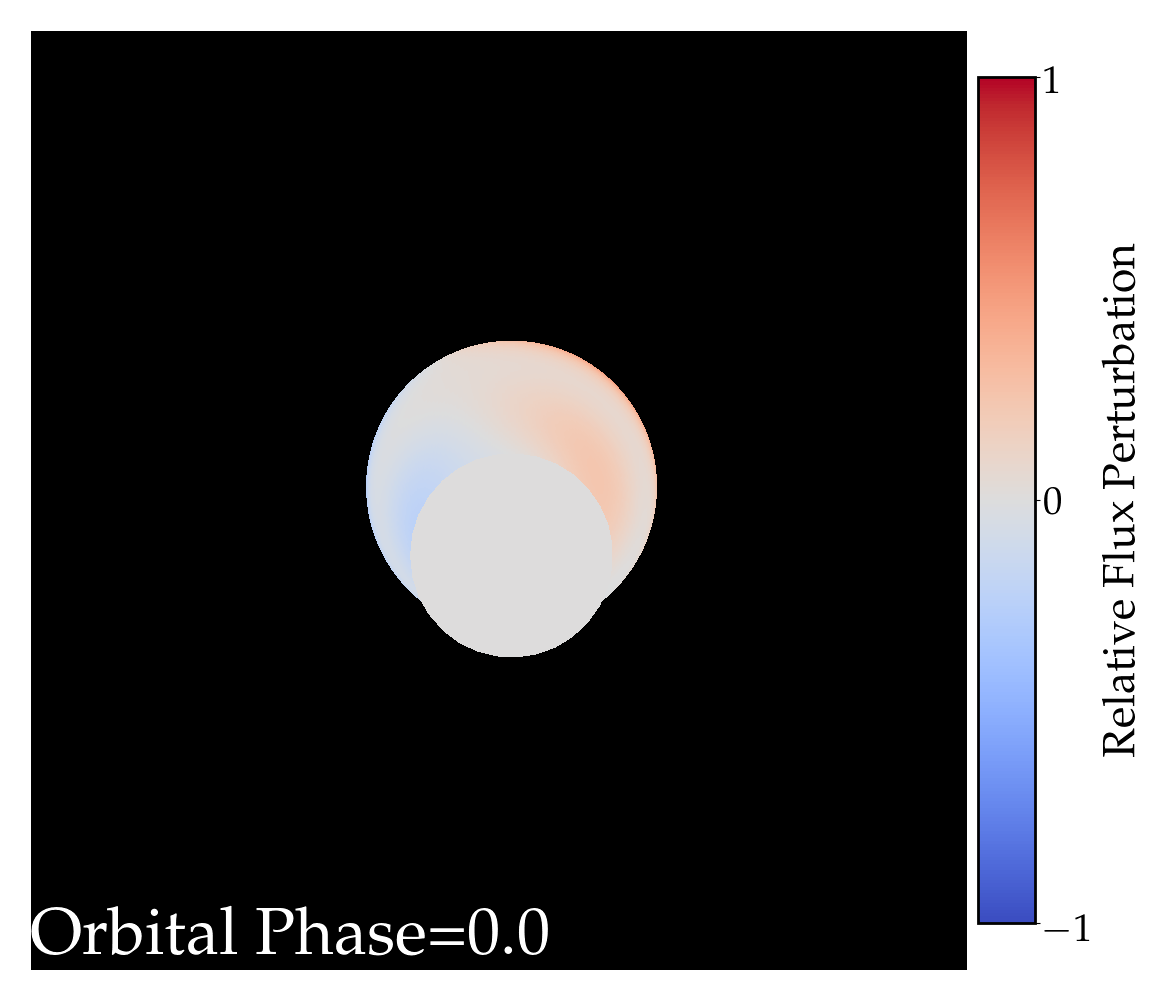}
\includegraphics[width=0.48\columnwidth]{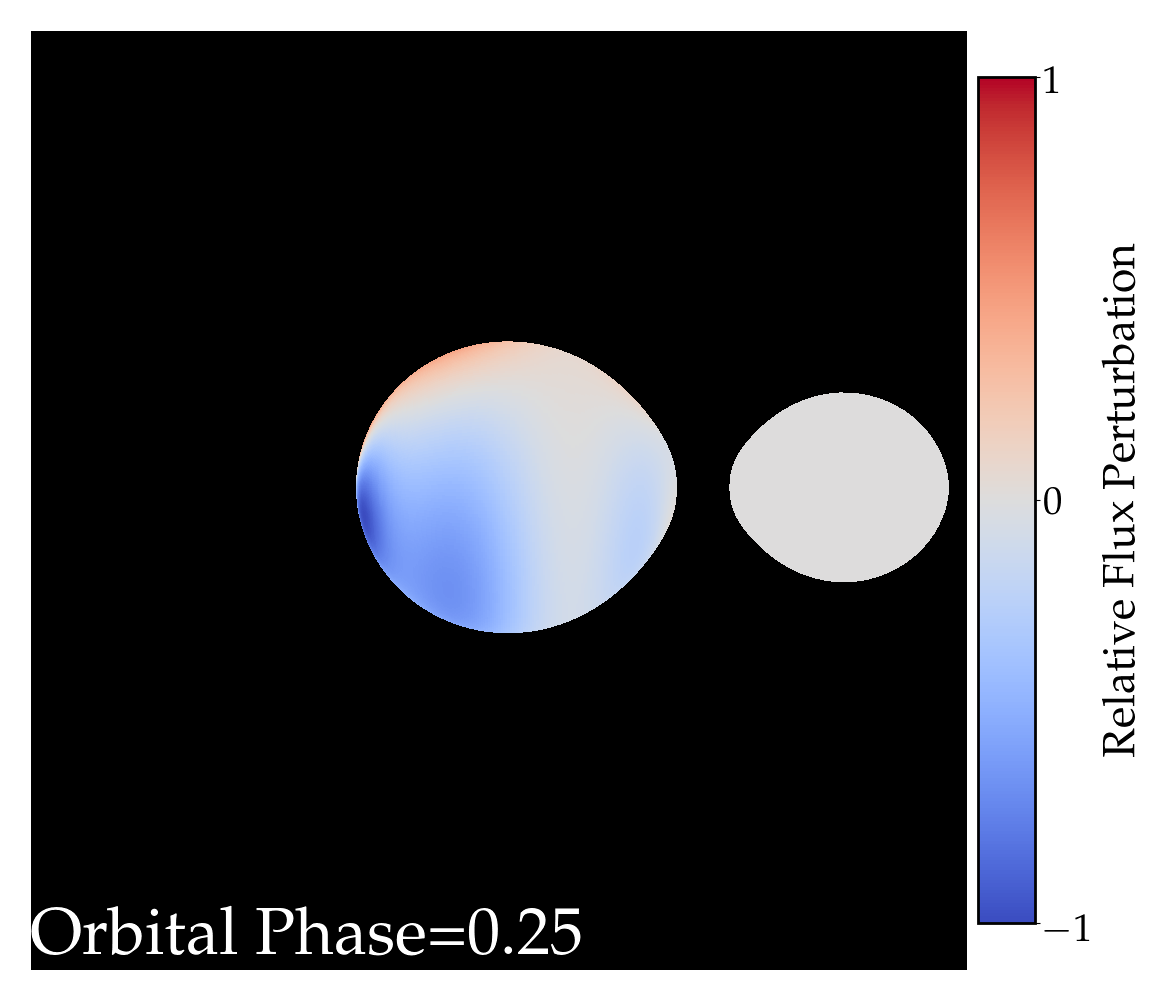}
\includegraphics[width=0.48\columnwidth]{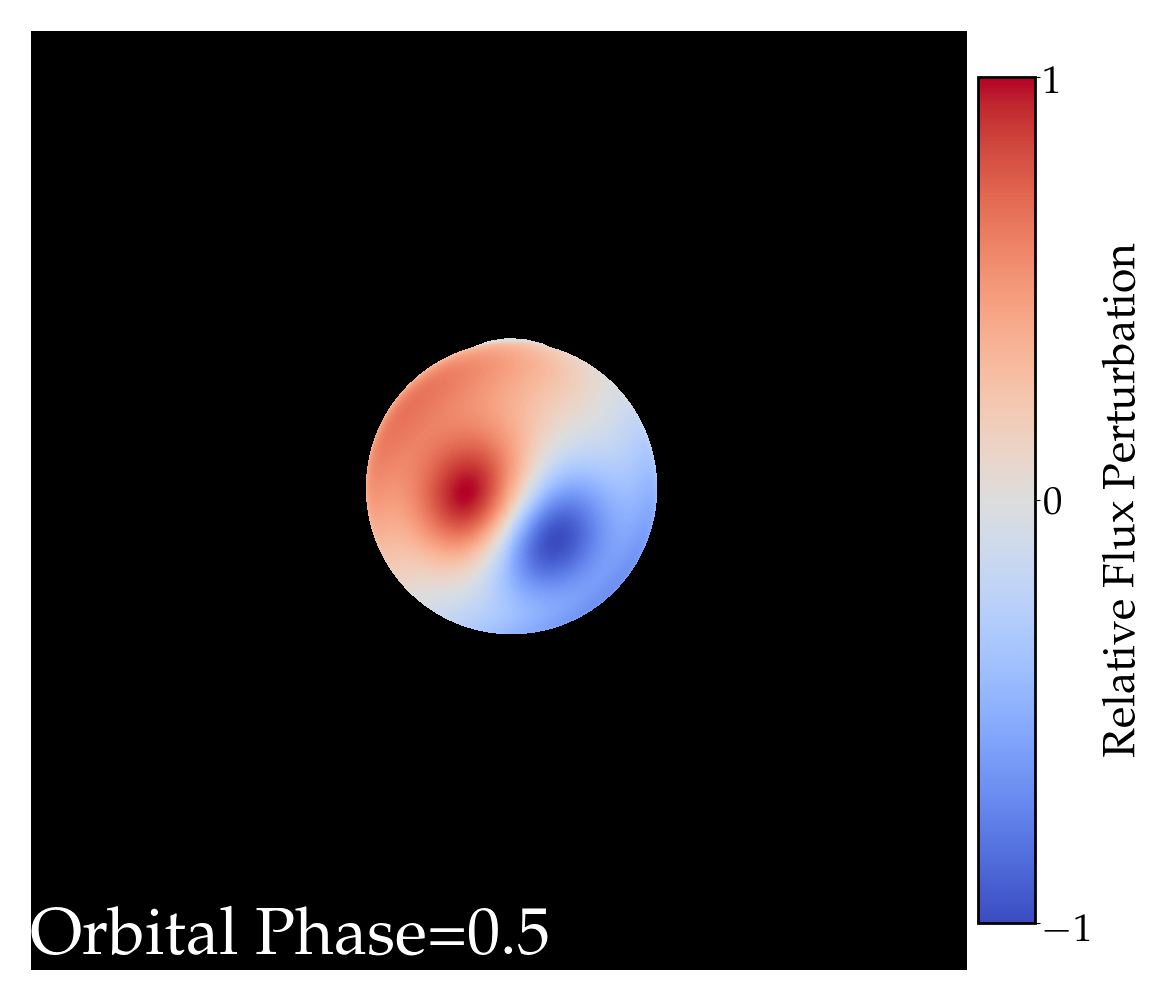}
\includegraphics[width=0.48\columnwidth]{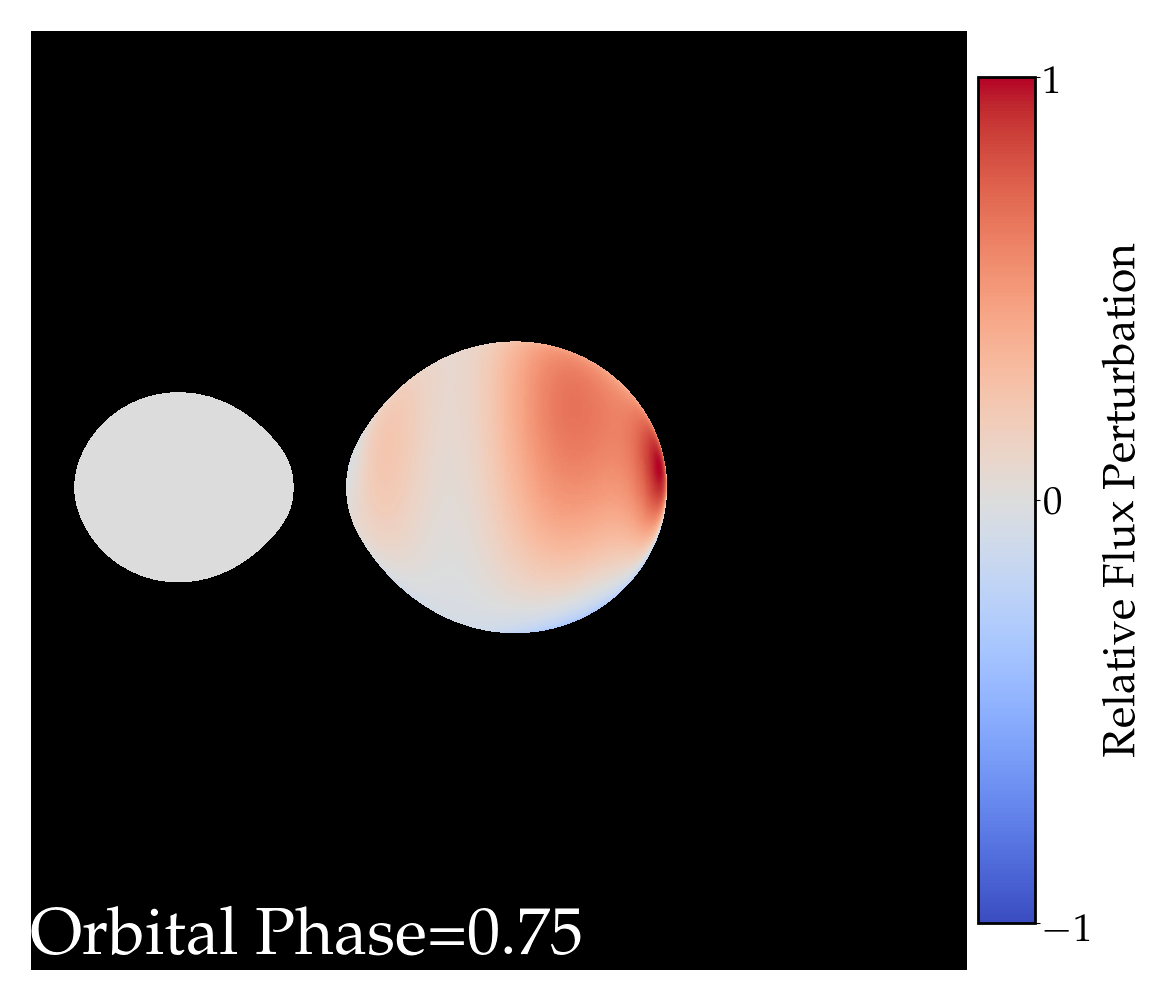}
\caption{\label{fig:TIC63fig}
A three dimensional model of the pulsation at $\nu = 21.22 \, {\rm d}^{-1}$ in TIC~63328020. The system is viewed from an inclination of $i=79$ degrees, with parameters from Table \ref{table1}, and is fixed on the center of mass of the pulsator. See the electronic version of this article for an animated version of this figure.
}
\end{figure}

Fig. \ref{fig:TIC63fig} shows a three-dimensional model of the pulsation at $\nu=21.22 \, {\rm d}^{-1}$ in TIC~63328020, illustrating the $m=-1$ nature of the mode and how its flux perturbations appear at different orbital phases.

Fig. \ref{TIC63flux} shows the flux perturbations as a function of tidal latitude for the four modes in Fig. \ref{TIC63}. In this case, the flux perturbation should be multiplied by $\cos (m \phi)$ to compute the flux variation at different tidal longitudes. Unlike normal $l=1$, $|m|=1$ modes, the flux perturbations are not largest at $\theta = \pi/2$, but rather are more oscillatory (due to their $\ell > 1$ components) and shifted towards smaller/larger tidal latitudes due to tidal trapping. The modes at $21.45 \, {\rm d}^{-1}$ and $18.51 \, {\rm d}^{-1}$ are largely trapped on the L$_1$ side, further suppressing their observed amplitude at orbital phase 0.5 in a manner inconsistent with the observations. The modes at $21.22 \, {\rm d}^{-1}$ and $18.23 \, {\rm d}^{-1}$ are not strongly trapped (though they do produce larger fluctuations on the L$_3$ side), so their amplitude modulation is smoother and more closely resembles the observations.

The modeled $m=-1$ modes at $21.22 \, {\rm d}^{-1}$ and $18.23 \, {\rm d}^{-1}$ also produce mode phase variations very similar to those observed, other than a nearly constant phase offset. However, note the sharp change in mode phase at orbital phase zero, which means that the observed phase can be shifted downwards by $\approx 2$\,rad if we change the zero point by only $\approx 0.02$ in orbital phase. Physically this would occur if the mode flux perturbation is shifted in orbital longitude by only $\approx 0.1$\,rad relative to the model. Allowing for this small offset produces a good match between the data and model.
The phase shifts of $\pi$ rad between orbital phases 0.25 and 0.75 is a natural consequence of the $m=-1$ mode geometry: at orbital phases 0.25 and 0.75, the mode produces a luminosity fluctuation that is a maximum on one side of the star, and a minimum on the other. This translates to an observed luminosity fluctuation that has the same amplitude at orbital phase 0.25 and 0.75, but which is shifted in phase by $\pi$\,rad. The observed modes are likely $m=-1$ modes as opposed to $m=1$ modes. While both would produce identical amplitude modulations, $m=1$ modes produce phase variations opposite to those of $m=-1$ modes, so the observed phases strongly suggest an $m=-1$ mode.

\section{Tidal Amplification}
\label{tidal}

A ``tidal amplification" effect may occur in the outer layers of the tidally distorted star. Because the density and sound speed drop sharply near the surface, the amplitude of a propagating acoustic wave increases, with the radial displacement, scaling as $\xi_r \propto (\rho r^2 c_s)^{-1/2}$ in the WKB limit. The surface flux perturbation of a mode scales as $\Delta F/F \propto \Delta T/T$, which approximately scales as $\Delta T/T \sim k_r \xi_r$ \citep{luan:17}. 
So, if pulsation energy is independent of latitude, we expect the mode visibility to scale approximately as 
\beq
\label{dff}
\frac{\Delta F}{F} \propto k_r \xi_r \propto (\rho r^2 c_s^3)^{-1/2}
\eeq
at the outer turning point of the wave. Because the acoustic cutoff frequency $\omega_c$ that determines the wave's outer turning point is dependent on tidal latitude, the mode's flux perturbation will vary with tidal latitude. The value of $\omega_c \sim g/c_s$ is smallest at the tidal poles where the effective gravity is smallest, allowing modes to propagate closer to the stellar surface to produce ``tidally amplified" flux perturbations.

\subsection{Polytrope-Roche modelling}
\label{poly}

\begin{figure}
\centering
\includegraphics[width=0.99\columnwidth]{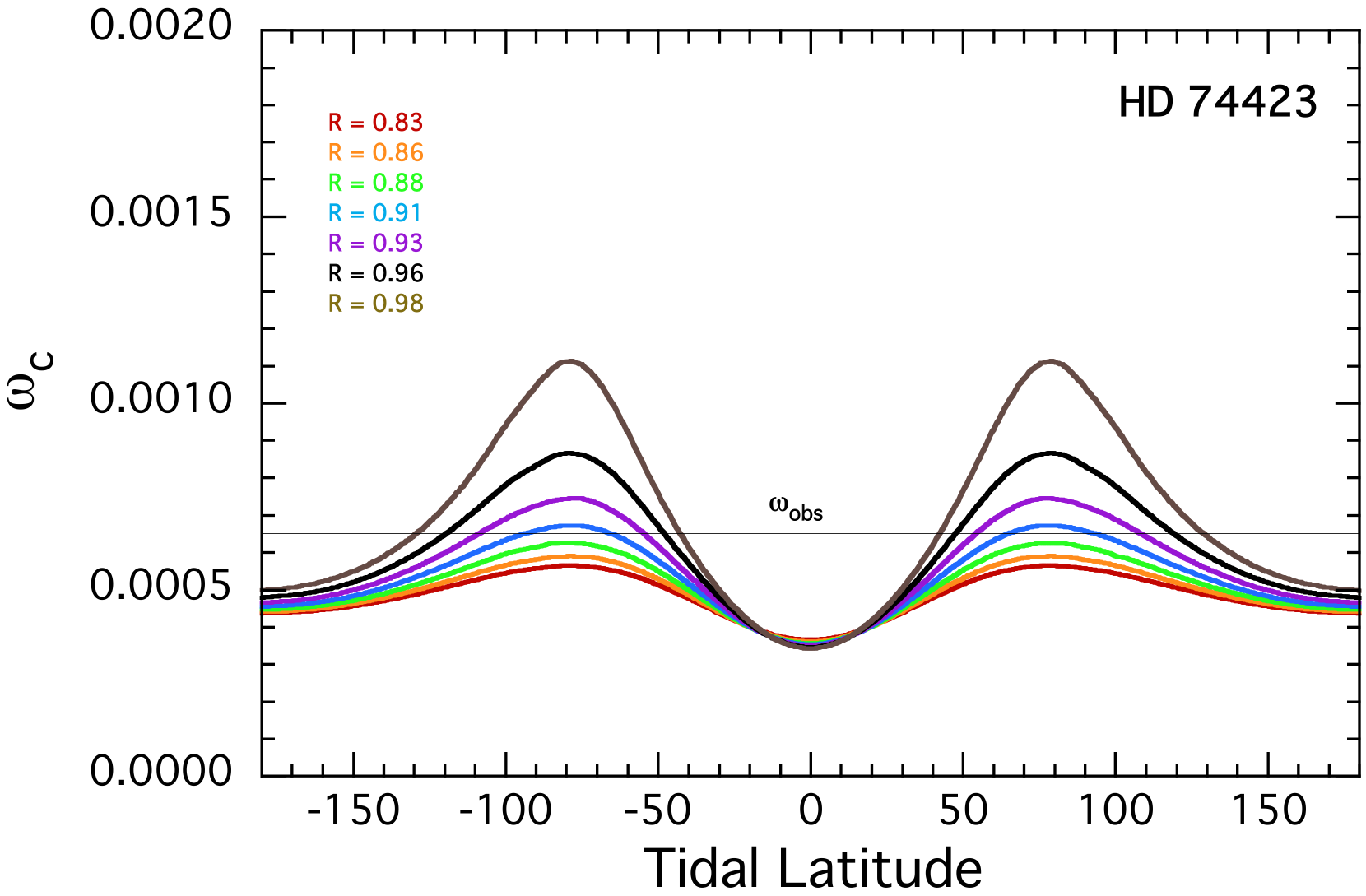} 
\caption{Acoustic cutoff frequency $\omega_c$ in the orbital plane, plotted as a function of tidal latitude, at several radial coordinates that are normalized in units of $R_y$. The horizontal gray line is the observed mode angular frequency $\omega_{\rm obs} = 6.4 \! \times \! 10^{-4}$ rad/s in HD~74423. Deep in the star, the observed mode propagates at all longitudes, but near the surface it only propagates near the L$_1$ point ($\theta=0$), producing a larger flux perturbation on that side of the star.}
\label{fig:azimuthal}
\end{figure}

To model the angular variations in the acoustic cutoff frequency $\omega_c$, we utilize a polytropic approximation to model the nearly Roche lobe-filling and distorted interior of a star like HD~74423. We describe this process in Appendix \ref{roche}.
We take the acoustic cutoff frequency as
\begin{equation}
\label{omc}
\omega_c \simeq \frac{c_s}{2H}  \simeq \frac{(1 + 1/n) g}{2 c_s} \, ,
\end{equation}
where $n$ is the polytropic index, and $c_s =\sqrt{\gamma P/\rho}$ is the  sound speed. The second equality in equation \ref{omc} stems from the relation $H=c_s^2/[(1+1/n)g]$ for polytropes. We adopt $n=3$ for our models, which is a good approximation for the surface layers of intermediate-mass stars.
Fig.~\ref{fig:azimuthal} shows corresponding plots of $\omega_c$ as a function of tidal latitude $\theta$ in the equatorial plane, for several different radial coordinates. Note that $\omega_c$ can be much smaller near the tidal pole at $\theta=0$, especially in the surface layers, allowing acoustic modes to propagate close to the photosphere to create large flux perturbations.

\begin{figure}
\centering
\includegraphics[width=0.99\columnwidth]{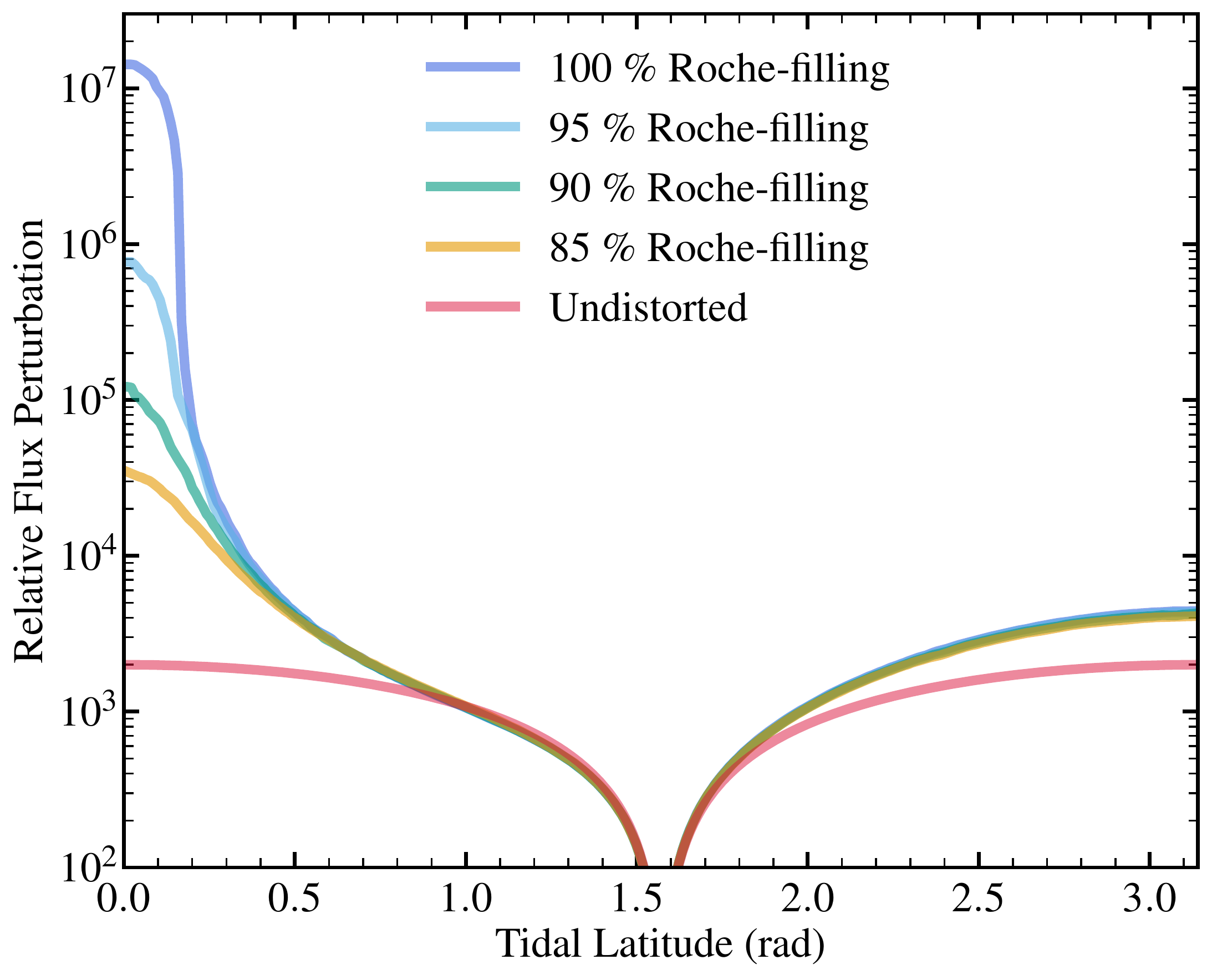} 
\caption{Relative flux perturbation as a function of tidal latitude at the mode's outer turning point, for models of HD~74423 with different Roche filling factors.}
\label{fig:dF}
\end{figure}

Using the acoustic cutoff frequency from our tidally distorted polytropic models above, we can calculate the mode's outer turning point as a function of tidal latitude. We then use equation \ref{dff} to calculate the relative flux perturbation as a function of tidal latitude in the orbital plane of our polytropic model. Fig. \ref{fig:dF} shows this estimate of the relative mode flux perturbation as a function of tidal latitude for polytropic models of HD~74423, but varying Roche filling factors (i.e., different orbital separations). It is clear that the flux perturbation can be much larger near the L$_1$ point. While more detailed models and mode eigenfunction calculations (including non-adiabatic and non-WKB effects) are needed for robust estimates, these simple models indicate the large amount of tidal amplification that can occur near the L$_1$ point.

\begin{figure}
\centering
\includegraphics[width=0.99\columnwidth]{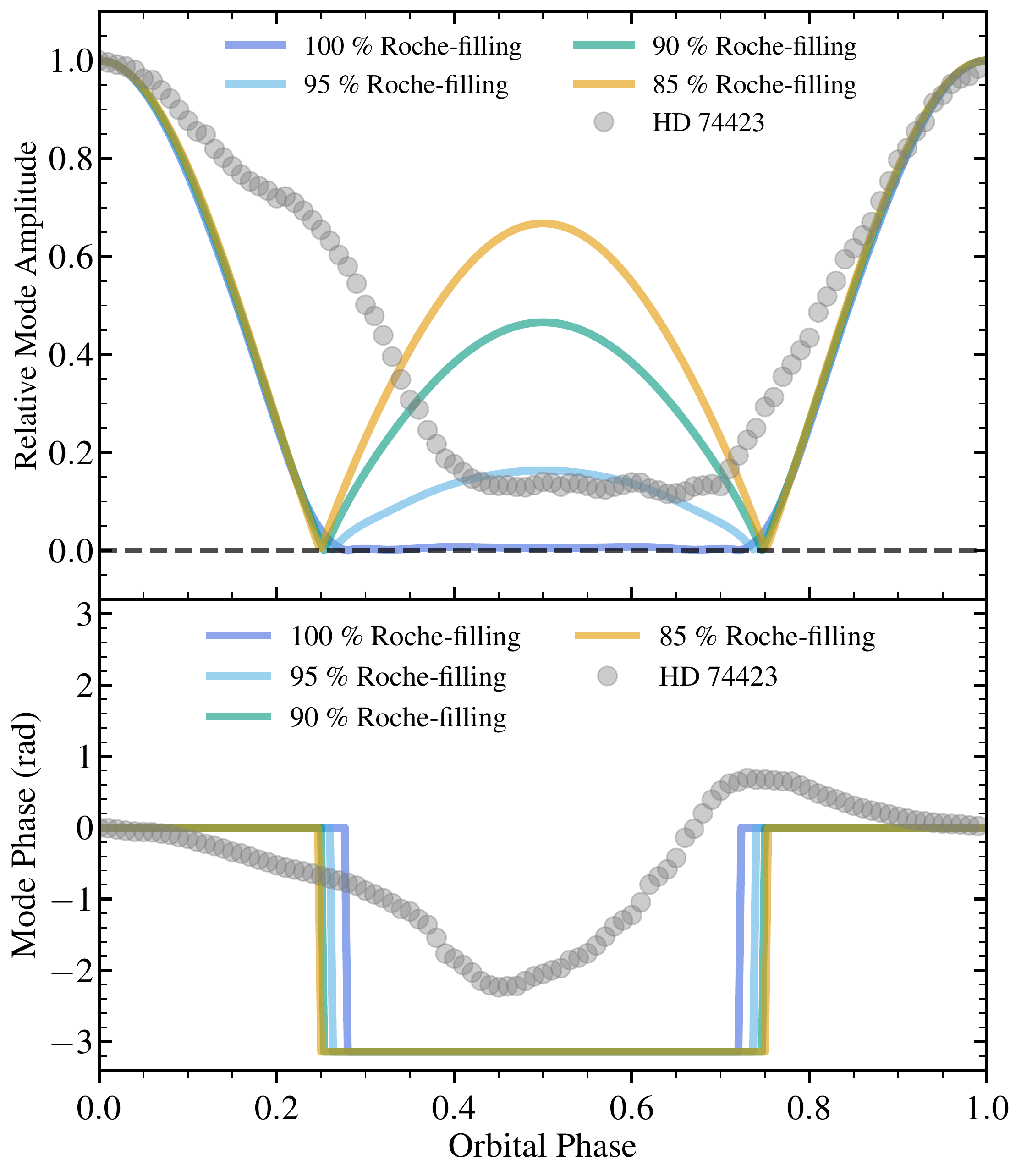} 
\caption{Observed and modeled amplitude and phase modulations of the pulsation mode in HD~74423, using our semi-analytic theory. This plot assumes the underlying mode is a dipole mode, and the different lines correspond to different Roche-filling factors of the primary star, whose local flux perturbations are shown in Fig. \ref{fig:dF}.  }
\label{fig:ampl0}
\end{figure}

Fig. \ref{fig:ampl0} shows the predicted amplitude and phase modulation of a dipole mode for tidally distorted models with relative flux perturbation amplitudes shown in Fig. \ref{fig:dF}. We see that the predicted amplitude modulation closely resembles the observed modulation in HD~74423 for large Roche filling factors. The peak amplitude occurs when the L$_1$ side of the star faces toward the observer, because of the large flux perturbation on that side. The Roche-filling models are also successful in producing the flat-bottomed minima, which occur when the L1 point is occulted behind the star. As the Roche filling factor decreases, the relative mode amplitude at orbital phase 0.5 increases due to the smaller flux perturbation at the L$_1$ point. The bottom panel of Fig. \ref{fig:ampl0} shows the observed and predicted pulsation phase modulation throughout the orbit, showing a phase shift of $\pi$ due to the dipole nature of the mode.

The analytic theory presented in this section may provide some intuition for why tidal trapping can occur on the L$_1$ side of the star. In the language of Section \ref{formalism}, such trapping occurs due to coupling with modes of different $\ell$, and coupling with modes of higher frequency $\omega$ which produce larger flux perturbations. However, the analytic theory fails in many respects. First, it cannot explain the large tidal trapping of the modes in CO~Cam, which is far from Roche-filling. Second, it cannot account for modes that are trapped away from the L$_1$ side of the star, as it always predicts larger flux perturbations on the L$_1$ side.

We suspect the main reasons for these failures are the WKB approximation inherent to the theory. For modes of low angular number $\ell$ interacting with the tidal distortion (which itself is dominated by $\ell=2$ and $\ell=3$), a WKB theory is insufficient, and more detailed calculations like those of Section \ref{formalism} are required. \cite{springer2013} also presents a useful WKB analysis of pulsations of tidally distorted stars, which focuses on modes with very high frequency and angular wavenumber such that the WKB approximation is more appropriate. They also find increased acoustic energy at tidal latitudes near zero, and surprisingly, increased acoustic energy near tidal latitudes of $\approx \! 130$ degrees, at least for the one angular wavenumber ($\ell \approx 50$) they consider.  Unfortunately, such high $\ell$ modes are very difficult to detect and unlikely to be observed in the systems discussed here.

\section{Discussion}
\label{disc}

\subsection{Limitations of This Work}

The biggest limitation of this paper is our neglect of Coriolis and centrifugal forces on the mode dynamics. The centrifugal and tidal distortion have similar amplitude in the quadrupolar surface displacement that they produce (see \citealt{preece2019}), with ratio
\begin{equation}
\label{tideratio}
\frac{\Delta R_{\rm tide}}{\Delta R_{\rm cen}} = \frac{3 M_2}{M_1 + M_2} \, .
\end{equation}
Tidal distortion is larger when $M_2 > M_1/2$, though the two have similar amplitude unless $M_2 \ll M_1$. Table \ref{table1} shows that $\Delta R_{\rm tide}/\Delta R_{\rm cen} \sim 1$ for the three systems considered here. Though we have included the components of the centrifugal force that are axisymmetric about the tidal axis, the non-axisymmetric components have similar magnitude. Both the non-axisymmetric component of the star's centrifugal distortion and the Coriolis force will induce coupling between modes of different $m$ about the tidal axis. If we had adopted the rotational axis for our coordinate system, then it would be the tidal force that induces non-axisymmetric couplings. This coupling can be handled by our method, but it greatly complicates the problem because it requires modes of all values of $m$ to be computed simultaneously, which greatly increases the size of the matrices (equation \ref{TV}) that need to be inverted.

Non-axisymmetric coupling will also complicate interpretation because the modes will no longer have an axis of symmetry. It is likely that all sorts of modes can exist in this situation, i.e., modes trapped at either tidal pole or the tidal equator, in addition to modes trapped near the rotational axis or equator, as well as modes trapped somewhere in between. In general, tidal modulation of the mode amplitude and phase will occur whenever modes of different $m$, measured in the frame aligned with the rotation/orbital axis, contribute to the observed pulsation amplitude and phase. 

Accounting for the full centrifugal and Coriolis forces will likely increase the level of mode mixing and avoided crossings that occur in a star, because there will be a much larger set of coupled modes within the same frequency range. Hence, including these effects is likely to produce more modes that are trapped in certain parts of the star. It will also likely increase the number of mode clusters like that seen in CO~Cam composed of a group of g~modes in an avoided crossing with a p~mode or f~mode. The increased coupling may also increase the number of modes involved in these clusters, allowing our model of CO~Cam (which contains only three observable trapped modes) to better reproduce the four modes observed in that system. Hence, these forces will need to be included in future work to get a full picture of the diversity of modes that occur in tidally distorted stars.

While often not very important for acoustic modes, the Coriolis force does play an important role in symmetry breaking. Consider the frame aligned with the rotation axis. 
The centrifugal and tidal forces only induce mixing proportional to $m^2$, whereas Coriolis forces induce mixing proportional to $m$, breaking the symmetry with respect to orbital phase. Such asymmetry is needed to explain the observations of HD~74423 (Fig. \ref{TIC35}), whose observed amplitude and phase variations are markedly asymmetric with respect to orbital phase. We suspect the asymmetry is most prominent for HD~74423 because it has the highest ratio of spin frequency to mode frequency ($\Omega/\omega = 0.07$) of the systems considered in this work.

\subsection{Tidal Alignment}

In this paper, we have implicitly assumed (through our neglect of Coriolis and centrifugal forces) that modes are tidally aligned, i.e., the tidal axis is their axis of symmetry. In reality, there is a competition between Coriolis forces and centrifugal distortion that promote rotational alignment, and tidal distortion that promotes tidal alignment. As with magnetically tilted modes in roAp stars \citep{Bigot2002}, tidally tilted modes will not be completely aligned with either axis, and we suggest that a mode's self coupling coefficients can be used to determine whether rotational or tidal alignment is preferred. 

To do this, one can evaluate the diagonal components of the matrices in equation \ref{TV} (i.e., the self-coupling terms) due to tides, and compare them with similar terms due to centrifugal distortion and the Coriolis force. For the Coriolis force, the relevant term is the familiar rotational splitting coefficient
\beq
\label{dC}
c_\alpha = \frac{\Omega \int \rho r^2 (2 U V + V^2) dr}{\omega \int \rho r^2 (U^2 + \ell(\ell+1) V^2 dr} \, .
\eeq
This should be compared to the similar terms from equations \ref{dT} and \ref{dV} from appendix \ref{math}, normalizing by mode inertia as in equation \ref{dC}.

For the systems in this work, we find that modes of primarily g~mode character have larger Coriolis coupling coefficients, while modes of primarily p~mode character have larger tidal coupling coefficients. The main reason is that the integrand of the numerator of equation \ref{dC} is small for p~modes, which reflects the usual insensitivity of p~modes to Coriolis forces. This is in addition to the fact that $c_\alpha$ is smaller for higher frequency modes due to the factor of $\Omega/\omega$.

The competition between tidal and centrifugal distortion is simpler because they both depend on the same coupling coefficients from equations \ref{dT} and \ref{dV}. The ratio of tidal to centrifugal terms is simply the ratio of tidal ellipticity to centrifugal ellipticity from equation \ref{tideratio}. Assuming synchronized rotation, tidal alignment is more likely for higher mass companions, and rotational alignment is more likely for lower mass companions.

We conclude that tidal alignment is likely to occur for p~modes in the binaries discussed in this paper. For stars in wider binaries, the tidal force rapidly diminishes in strength relative to the Coriolis force, and tidal alignment is much less likely. Assuming spin-orbit synchronism of systems similar to those examined in this paper, tidally aligned p~modes could exist out to orbital periods of $\sim 10\,{\rm days}$. However, $\delta$~Sct stars typically rotate faster than synchronous for binaries wider than several days, so we expect tidal alignment to be most common in binaries with periods less than a few days. Higher frequency modes could remain tidally aligned out to longer orbital periods. It is also possible that some systems could simultaneously exhibit both tidally aligned modes and rotationally aligned modes. We do not expect g~modes to be tidally aligned, except for hybrid g~modes that have significant p~mode character, or perhaps for modes in some systems spinning slower than synchronously.

\subsection{Mode Selection and Non-adiabatic Effects}

In this work, we do not address the question of mode selection, i.e., which modes are excited to observable amplitudes. This problem is not understood even for spherical stars and is beyond the scope of this work. For each of our stellar models discussed in Section \ref{results}, we find unstable radial or dipole p~modes, but their radial orders are larger than those observed for HD 74423 and TIC 63328020. The disagreement could arise from problems in our spherical models, the neglect of turbulent pressure in mode driving \citep{antoci:14,xiong:16}, or it could result from tidal effects. Since lower temperature $\delta$~Sct stars typically exhibit lower frequency (and lower overtone) pulsations, the low-temperature gravity-darkened L$_1$ side of a tidally distorted star would likely result in stronger driving of low-frequency modes. Our models to not include latitudinal variations in temperature, nor do they account for tidal coupling when computing mode driving/damping rates, so they do not capture this effect.

Our models cannot reliably predict the $m$ values of tidally aligned modes excited in $\delta$~Sct stars. Like normal stars, modes of many values of $m$ can likely be excited by the $\kappa$-mechanism, and there is not an obvious reason why the tidal distortion (or centrifugal forces) should selectively excite any particular value of $m$. We found that both $m=0$ and $m=-1$ modes are necessary to explain the systems examined in this work, and we expect that other values of $m$ will be observed as more tidally tilted pulsators are discovered. Given the latitudinal variation in temperature of tidally distorted stars, it is possible that certain values of $m$, or modes tidally trapped on one side of the star, will have larger driving rates and will be more likely to be observed. This possibility should be investigated in future work.

\subsection{Predictions}

Tidally tilted pulsators may provide new opportunities for asteroseismology. The phase and amplitude modulation of modes in these systems provide extra information that allow for mode identification, though comparisons with models is complicated by the tidal mode coupling calculation that must be performed. Additionally, the pulsation pattern for low-frequency p~modes or f~modes is very sensitive to the stellar model, due to tidal coupling with core g~modes. In principle, this sensitivity may allow for very tight constraints on the stellar structure, provided that tidal coupling is properly accounted for, including the full effects of the Coriolis and centrifugal forces.

Another prediction of our models is that tidally tilted modes will exhibit diverse patterns of amplitude modulation. Considering just $m=0$ modes, the modes at $\nu = 9.9 \, {\rm d}^{-1}$ and $\nu = 9.2 \, {\rm d}^{-1}$ in Fig. \ref{TIC35} exhibit somewhat different amplitude modulation than the observed modes. We also find a nearly equal number of modes in the models that are trapped on the L$_3$ side of the star, such that their amplitudes would peak at orbital phase 0.5. Other modes can be trapped at mid-latitudes, such that their amplitudes would peak at orbital phases 0.25 and 0.75. Unless mode excitation effects prevent such modes from being excited, we expect tidally aligned pulsations to exhibit many different and unique patterns of amplitude modulation in each system. 

For non-axisymmetric modes, the amplitude modulation can also look quite different. The mode with $\nu=18.51 \, {\rm d}^{-1}$ in Fig. \ref{TIC63} demonstrates some of the more complex behavior that can occur. In this case, tidal trapping on the L$_1$ side of the star creates an amplitude minimum near orbital phase 0.5, but with additional modulation due to the changing viewing geometry of the $m=-1$ mode pattern. Modes with $|m|=2$ could exhibit even faster variations in amplitude and phase over the orbit. The modulation patterns also depend on the orbital inclination of each system, further increasing the diversity. 

We suspect the tidally tilted pulsators discussed in this paper were the first three to be discovered because of their relatively simple amplitude modulation patterns and simple power spectra. HD~74423 exhibits only one pulsation mode that is obviously modulated in amplitude upon visual inspection of its light curve. This mode also produces one conspicuous multiplet in the star's power spectrum, which is easy to identify. TIC~63328020 also exhibits only one oscillation mode, whose amplitude rises and falls twice per orbit. CO~Cam exhibits several oscillation modes, but their amplitudes are all modulated in the same way, again making it easy to identify in a visual inspection of the light curve.

It is likely that there are many stars with multiple tidally tilted pulsations that are waiting to be discovered. Because each mode's amplitude could be modulated differently over the orbit, such stars could be challenging to identify from a visual inspection of their light curves. Their power spectra would also be very complex, with many overlapping multiplets (each corresponding to a tidally tilted mode) with peaks split by the orbital frequency. Discovering and measuring this sort of amplitude modulation will be more challenging, but may be achievable by finding power spectra with a large number of peaks split by exactly the same frequency (i.e., the orbital frequency), but which are not themselves orbital harmonics like the tidally excited oscillations in eccentric binaries. We predict many more tidally tilted pulsators will be observed upon detailed examination of the p~mode pulsations of stars in very close binaries.


\section{Conclusions}

We have conducted a detailed examination of the effect of tidal distortion on the pulsation modes of stars in close binary systems. Unlike most prior work that focused on tidal perturbations to mode frequencies, we have examined how tidal distortion affects mode eigenfunctions, creating tidally aligned and tidally trapped pulsation modes. The tidal asphericity bends and focuses the paths of waves propagating through the star, which can cause the resulting oscillation modes to be trapped within (or away from) the tidal bulges. We presented a formalism to compute the modes of tidally distorted stars by expanding in the basis of modes of spherically symmetric stars. Because the tidal asphericity couples modes of different angular numbers $\ell$, the pulsation modes of tidally distorted stars are superpositions of many values of $\ell$, creating eigenfunctions that can be localized to one region of the star, i.e., tidally trapped pulsations.

The observational manifestation of this tidal trapping is the newly discovered class of ``single-sided" and ``tidally tilted" pulsators in close binaries. These stars contain oscillation modes that are aligned with the tidal axis, so the pulsation mode amplitudes and phases are modulated over the orbital phase due to the observer's changing viewing geometry. We have applied our tidal coupling theory to stellar models of the tidally tilted pulsators HD~74423 \cite{handler2020}, CO~Cam \citep{kurtz2020}, and TIC~63328020 (Rappaport et al., in prep). Figs \ref{TIC35}, \ref{COCam}, and \ref{TIC63} show our main results. Our tidal trapping theory can largely explain the observed mode amplitude variations in each of these unique systems. HD~74423 contains a single acoustic mode strongly trapped on either the L$_1$ or L$_3$ side of the star, CO~Cam contains a cluster of tidally trapped hybrid fundamental/gravity modes (which results from tidal coupling between modes of different $\ell$), and TIC~63328020 contains a non-axisymmetric $m=-1$ tidally tilted mode.

The currently known tidally tilted pulsators exhibit a small number of pulsation modes, simplifying their power spectra and making them easier to identify. Stars with larger numbers of tidally tilted pulsations will be harder to identify but may provide new asteroseismic diagnostics for the structures of tidally distorted stars. Future theory should incorporate the Coriolis and centrifugal forces into models, which complicates the calculation but is necessary to fully capture the mode dynamics. It also remains unclear which types of tidally trapped pulsations are most likely to be excited to observable amplitudes. Given the recent surge in discoveries of tidally tilted pulsators, many new systems are likely to be uncovered in the near future, so more comprehensive models will be needed to solve the puzzles that are certain to arise.

\section*{Acknowledgments}

We thank the anonymous referee for a thorough review of this manuscript. This research was supported in part by the National Science Foundation under Grant No. NSF PHY-1748958. JF is thankful for support through an Innovator Grant from The Rose Hills Foundation, and the Sloan Foundation through grant FG-2018-10515. GH gratefully acknowledges funding through NCN grant 2015/18/A/ST9/00578.

\section*{Data Availability}
Data and source code is available upon request to the authors.

\bibliographystyle{mnras}
\bibliography{CoreRotBib,CoreRotBib_2}

\appendix

\section{Mode Coupling Coefficients}
\label{math}

To compute the coupling coefficients between modes of a tidally distorted star, we follow the calculation and terminology of \cite{dahlen1998}, described in Appendix D. The Woodhouse kernels can be calculated either in terms of the perturbed density, pressure, etc., or in terms of the stellar ellipticity. The latter option is much better for stars, because a linear Eulerian description of the perturbed stellar structure breaks down if the tidal distortion is larger than a scale height, as it is near the surfaces of stars. Each component of the tidal ellipticity is
\beq
\varepsilon = \sum_{\ell_t} \varepsilon_{\ell_t} \, 
\eeq
with 
\begin{align}
\label{epsl}
\frac{2}{3} \varepsilon_{\ell_t} &= \sqrt{\frac{2 \ell_t +1}{4 \pi}} \frac{U_{\ell_t}}{r g Y_{\ell_t 0}} \nonumber \\
&= -\frac{M_2}{m(r)} \bigg(\frac{r}{a}\bigg)^{\ell_t +1}
\end{align}
and the second line follows from the tidal potential as given by equation \ref{Utide}, with $m(r)$ and $r$ the mass and radius coordinates within the star before adding tidal distortion. Below we will also encounter the radial derivative of the tidal ellipticity,
\begin{align}
\eta &= \frac{\partial \ln \varepsilon_{\ell_t}}{\partial \ln r} \nonumber \\
&= \ell_t + 1 - \frac{4 \pi \rho r^3}{m(r)} \, .
\end{align}

From equation D.80 of \cite{dahlen1998}, the kinetic energy coupling coefficient between two modes, indexed by $\alpha$ (with spherical harmonic $\ell$ and $m$) and $\alpha'$ (with spherical harmonic $\ell'$ and $m'$) is
\beq
\label{dT}
\delta T_{\alpha \alpha'} = \int^R_0 \frac{2}{3} \varepsilon \rho r^2 \big[ \bar{T}_\rho - (\eta + 3) \check{T}_\rho \big] dr 
\eeq
with 
\beq
\label{dT2}
\bar{T}_\rho = -Z_{\ell \ell' \ell_t}^{m m' m_t} U V' - Z_{\ell' \ell \ell_t}^{m' m m_t} U' V 
\eeq
and
\beq
\label{dT3}
\check{T}_\rho = X_{\ell \ell' \ell_t}^{m m' m_t} U U' + Z_{\ell \ell_t \ell'}^{m m_t m'} V V' \, .
\eeq
In these expressions, $U$ is the radial displacement associated with mode $\alpha$, and $U'$ is the radial displacement for mode $\alpha'$. Similarly, $V$ is the horizontal displacement. We have already factored out the time and angular dependence, i.e., the full displacement is $\bxi = U(r) Y_{\ell m}(\theta,\phi) e^{-i \omega t} + V(r) r\nabla_\perp Y_{\ell m}(\theta,\phi) e^{-i \omega t}$, such that $U$ and $V$ are functions only of $r$.

In equations \ref{dT2}-\ref{dT3}, $X_{\ell_1 \ell_2 \ell_3}^{m_1 m_2 m_3}$ and $Z_{\ell_1 \ell_2 \ell_3}^{m_1 m_2 m_3}$ are angular overlap integrals between mode $\alpha$, mode $\alpha'$, and the component of the tidal potential with $\ell = \ell_t$ and $m = m_t = 0$ in our coordinate system. Explicitly, 
\begin{align}
X_{\ell \ell_t \ell'}^{m m_t m'} & = \bigg(\frac{4 \pi}{2 \ell_t +1}\bigg)^{1/2} \int d S Y_{\ell m}^* Y_{\ell_t m_t} Y_{\ell' m'} \nonumber \\
& = (-1)^m \big[(2 \ell + 1)(2 \ell' + 1)\big]^{1/2} \nonumber \\
& \times \begin{pmatrix}
\ell & \ell_t & \ell' \\
-m & m_t & m'
\end{pmatrix}
\begin{pmatrix}
\ell & \ell_t & \ell'\\
0 & 0 & 0
\end{pmatrix}
\end{align}
where the terms in parentheses are Wigner 3-j symbols. Additionally, 
\begin{align}
\label{z}
Z_{\ell \ell_t \ell'}^{m m_t m'} &= r^2 \int d S \, Y_{\ell_t m_t} \nabla_\perp \! Y_{\ell m}^* \cdot \nabla_\perp \! Y_{\ell' m'} \nonumber \\
&= \frac{1}{2} \big[ \ell(\ell+1) + \ell'(\ell'+1) - \ell_t(\ell_t+1) \big] X_{\ell \ell_t \ell'}^{m m_t m'}
\end{align}
We will drop the $\ell$ and $m$ subscripts and superscripts from $X$, which is invariant in exchanges between modes. For $Z$, we use the shorthand $Z = Z_{\ell_t \ell \ell'}^{m_t m m'}$, $Z' = Z_{\ell \ell' \ell_t }^{m m' m_t}$, and $Z_t = Z_{\ell' \ell_t \ell}^{m' m_t m}$, i.e., the symbol denotes which value of $\ell$ accounts for the negative term in equation \ref{z}. 

The potential energy coupling terms are
\beq
\label{dV}
\delta V_{\alpha \alpha'} = \int^R_0 \frac{2}{3} \varepsilon r^2 \bigg( \kappa \big[\bar{V}_\kappa - (\eta +1) \check{V}_\kappa \big] + \rho \big[ \bar{V}_\rho - (\eta + 3) \check{V}_\rho \big] \bigg) dr 
\eeq
and the incompressibility is $\kappa = \rho c_s^2 = \Gamma_1 p$, where $c_s$ is the sound speed and $p$ is the pressure. The integrand components are 
\begin{align}
\bar{V}_\kappa &= - X d U/d r (d U'/d r + f') - X d U'/d r (d U/d r + f) \nonumber \\
& - Z' V (d U'/d r + f')/r - Z V' (d U/d r + f )/r
\end{align}
\begin{align}
\check{V}_\kappa &= \frac{1}{2} X (-d U/d r +f)(d U'/d r + f') \nonumber \\
&+ \frac{1}{2} X (-d U'/d r + f')(d U/d r + f) \nonumber \\
& + Z' V (d U'/d r + f')/r + Z V' (d U/d r + f)/r
\end{align}
\begin{align}
\bar{V}_\rho &= X (r d P/d r + 4 \pi G \rho r U + g U) f' \nonumber \\
& + X (r d P'/d r + 4 \pi G \rho r U' + g U') f \nonumber \\
& - Z' g V U'/r - Z g V' U/r + 3 X g U U'/r + 3 X g U' U/r \nonumber \\
& + Z_t P V'/r + Z_t P' V/r \nonumber \\
&- \ell(\ell+1) X P U'/r - \ell'(\ell'+1) X P' U/r 
\end{align}
\begin{align}
\check{V}_\rho &= X U d P'/d r + X U' d P/d r + 4 \pi X G \rho U U' + 4 \pi X G \rho U' U \nonumber \\
& - Z g U V'/r - Z' g U' V/r + Z_t V P'/r + Z_t V'P/r 
\end{align}
Here, $P$ is the the Eulerian gravitational potential perturbation $\delta \Phi = P Y_{\ell m} e^{- i \omega t}$, and we define $f = \big[2 U - \ell (\ell+1) V \big]/r$.

The full expressions include additional terms including the toroidal displacement $W$, which we have ignored because $W=0$ for poloidal modes of non-rotating stars. This is acceptable for our purposes, but including toroidal components is very important for low-frequency gravito-inertial modes. Technically, there are also additional terms that arise from the perturbed gravity field of the star, the $V_\Phi$ terms from \cite{dahlen1998}. However, these terms disappear when combining the perturbed gravitational field and the tidal field, so they are not reproduced here. Note that the coupling coefficients are identical under the exchange of $\alpha$ and $\alpha'$, which ensures that the matrices in equation \ref{TVnew} are symmetric. The potential and kinetic energy operators are Hermetian (when using adiabatic mode eigenfunctions) such that the eigenvalues remain real. This ensures the mode frequencies $\omega$ are either purely real (stable) or purely imaginary (unstable). We verify that all of the perturbed eigenfrequencies of our models remain stable.

In our calculations, we include the $\ell=2$ component of the centrifugal distortion that is axisymmetric in the tidal frame. We assume spin-orbit synchronization such that the angular rotation frequency is $\Omega^2 = G (M_1+M_2)/a^3$. One can show that the associated ellipticity of this component is,
\beq
\varepsilon_{\rm cen} (\ell_t=2,m_t=0) = \frac{1}{6} \frac{M_1+M_2}{M_1} \varepsilon (\ell_t=2,m_t=0) \, 
\eeq
with $\varepsilon (\ell_t=2,m_t=0)$ evaluated from equation \ref{epsl}. Hence, we increase the value of $\varepsilon (\ell_t=2,m_t=0)$ by a factor of $1+(M_1+M_2)/(6M_1)$ to account for this component of the centrifugal distortion. The $\ell=0$ component of the centrifugal distortion does not mix modes of different $\ell$, but it does mix modes of the same $\ell$. It can be accounted for with a centrifugal component of the potential energy coupling to be added to equation \ref{dV},
\begin{align}
\label{dVcen}
\delta V_{\rm cen} &= \frac{2}{3} \Omega^2 \delta_{\alpha \alpha'} - \frac{2}{3} \Omega^2 \ell(\ell+1) \delta_{\ell \ell'} \nonumber \\
& \times \int^R_0 \rho r^2 \big( V V' + U V' + V U'\big) dr \, ,
\end{align}
where the first term is non-zero only for a mode coupling with itself.

\section{Observed mode amplitudes and phases}
\label{nonaxisymmetric}

Here we compute the amplitude and phase variation of a mode that is non-axisymmetric about the tidal axis, i.e., it has  $m\neq0$. This situation is a little more complicated because the system is no longer symmetric about the tidal axis and a decomposition into Legendre polynomials is no longer possible. 

We begin by decomposing the flux perturbation into spherical harmonics as in equation \ref{dfdecomp}, but this time allowing for non-axisymmetric spherical harmonics:
\begin{equation}
    \label{dfdecomp2}
    \Delta F(\theta,\phi) = \sum_\ell \Delta F_\ell  Y_{\ell m}(\theta,\phi) \, .
\end{equation}
Note that the sum only goes over $\ell$ because only modes of a single value of $m$ contribute as long as the system is symmetric about the tidal axis. Each component $\Delta F_\ell$ of the decomposition is
\begin{equation}
    \label{dfk}
    \Delta F_\ell = \int dS \,  \Delta F(\theta,\phi) Y_{\ell m}^*(\theta,\phi) \, ,
\end{equation}
where the integral is taken over a spherical surface, $dS = \sin \theta d\theta d \phi$.

If the function $\Delta F(\theta,\phi)$ is computed numerically from our method in Section \ref{tidal}, one must take a numerical overlap integral or Legendre polynomial decomposition to solve for each value of $\Delta F_\ell$. For our method in Section \ref{formalism}, $\Delta F(\theta,\phi)$ has already been decomposed into spherical harmonics such that
\begin{align}
\label{delf}
\Delta F(\theta,\phi) &= \sum_\alpha a_\alpha \Delta F_\alpha Y_{\alpha}(\theta,\phi) \nonumber \\
&= \sum_\ell \sum_n a_{\ell,n} \Delta F_{\ell,n} Y_{\ell m}(\theta,\phi) \, .
\end{align}
where the complex value of the surface flux perturbation of each basis mode, $\Delta F_\alpha$ is computed from the surface temperature perturbation $\Delta T_\alpha$ as described in Section \ref{application}. Here, $n$ is an index for each basis mode with angular number $\ell$. Inserting equation \ref{delf} into equation \ref{dfk}, we find
\beq
\Delta F_\ell = \sum_n a_{\ell,n} \Delta F_{\ell,n} \, ,
\eeq
i.e., it is the weighted sum of the surface flux perturbations of each basis mode with of angular number $\ell$.

The next step is to compute the observed flux variation in the observer's frame. Following the same procedure as Section \ref{amp}, we decompose each spherical harmonic in the tidal axis frame into spherical harmonics in the frame aligned with the orbital axis:
\begin{align}
    Y_{\ell m}(\theta,\phi) &= \sum_{m_{\rm s}=-\ell}^{\ell} D^\ell_{m_{\rm s},m}(\alpha,\beta,\gamma) Y_{\ell m_{\rm s}}(\theta_{\rm s},\phi_{\rm s}) \nonumber \\
    &= \sum_{m_{\rm s}=-\ell}^{\ell} d^\ell_{m_{\rm s},m}(-\pi/2) Y_{\ell m_{\rm s}}(\theta_{\rm s},\phi_{\rm s}) \, .
\end{align}
Here, $(\theta_{\rm s},\phi_{\rm s})$ are angular coordinates in the corotating frame aligned with the orbital axis, and $D$ is a Wigner function of the Euler angles $\alpha$, $\beta$, and $\gamma$. In this case, we can define the coordinate systems such that $\alpha=\gamma=0$, such that the Wigner D function reduces to the Wigner small $d-$matrix element $d^\ell_{m_{\rm s},m}(\beta)$, and $\beta = -\pi/2$ for the 90 degree rotation between the tidal and orbital axes. Performing this transformation, the flux variation across the stellar surface is 
\begin{equation}
    \label{dfdecomp3}
    \Delta F = \sum_\ell \Delta F_\ell \sum_{m_{\rm s}=-\ell}^{\ell} d^\ell_{m_{\rm s},m}(-\pi/2) Y_{\ell m_{\rm s}}(\theta_{\rm s},\phi_{\rm s})  \, .
\end{equation}

To compute the flux perturbation in the observer's frame, we again perform the rotation
\begin{align}
    Y_{\ell m_{\rm s}}(\theta_{\rm s},\phi_{\rm s}) &= \sum_{m_{\rm o}=-\ell}^{\ell} D^\ell_{m_{\rm o},m_{\rm s}}(\alpha_{\rm o},\beta_{\rm o},\gamma_{\rm o}) Y_{\ell m_{\rm o}}(\theta_{\rm o},\phi_{\rm o}) \nonumber \\
    &= \sum_{m_{\rm o}=-\ell}^{\ell} d^\ell_{m_{\rm o},m_{\rm s}}(i_{\rm o}) e^{-i m_{\rm s} \Omega t} Y_{\ell m_{\rm o}}(\theta_{\rm o},\phi_{\rm o}) \, .
\end{align}
Here, $\theta_{\rm o}$ and $\phi_{\rm o}$ are angular coordinates in the observer's frame, in which the observer is located at $\theta_{\rm o} = 0$ and the companion star is at $\phi_{\rm o}=\phi_{\rm s}=0$ at time $t=0$. For this coordinate transformation, the angle $\gamma_{\rm o} = \Omega t$ is the orbital phase of the pulsating star, the angle $\beta_{\rm o} = i_{\rm o}$ is the observer's inclination, and $\alpha_{\rm o}=0$. Upon integrating over the surface of the star to obtain the observed luminosity variation $\Delta L$, only the $m_{\rm o}=0$ term remains, and a limb-darkening coefficient $b_\ell$ (see equation \ref{bl}) appears. We then obtain
\begin{align}
\Delta L &= 2 \pi \sum_\ell \sqrt{\frac{2\ell +1}{4 \pi}} \,  b_\ell \, \Delta F_\ell \nonumber \\
& \times \sum_{m_{\rm s}=-\ell}^\ell d^\ell_{m_{\rm s},m}(-\pi/2) d^\ell_{0,m_{\rm s}}(i_{\rm o}) e^{-i m_{\rm s} \Omega t} \, .
\end{align}
The summation is easily computed, and we find including values up to $\ell=6$ is necessary to obtain converged light curve models. Note that the final luminosity amplitude and phase are a function of the mode's angular flux distribution, the inclination angle $i_{\rm o}$, and the time $t$.

\section{Roche Model of Tidally Distorted Star}
\label{roche}

The effective potential in a frame rotating with the binary orbit can be written as:
\begin{equation}
\Psi_{\rm eff} = \Psi_1 +\Psi_2+\Psi_{\rm cent}
\label{eqn:psi}
\end{equation}
where the three terms are the potential of star 1, the potential of star 2 (taken to be a point mass), and the fictitious centrifugal potential, respectively.  The equation of hydrostatic equilibrium in the rotating frame can be expressed as:
\begin{equation}
\frac{1}{\rho} \vec{\nabla}P = \vec{g}_{\rm eff} = -\vec{\nabla} \Psi_{\rm eff} \, ,
\label{eqn:grav}
\end{equation}
see Eqns. (1) and (8) of \cite{hachisu1986}. For a polytropic gas, $P = K \rho^{1+1/n}$, where $n$ is the polytropic index, and $K$ is the constant of proportionality. The left-hand side of Eqn.~(\ref{eqn:grav}) can be written for the case of a polytrope as 
\begin{equation}
\frac{1}{\rho} \vec{\nabla}P =K (n+1) \vec{\nabla} \rho^{1/n}.
\end{equation}
In that case we can rewrite the equation of hydrostatic equilibrium as:
\begin{equation}
\vec{\nabla} \left[K (n+1) \rho^{1/n} +\Psi_{\rm eff} \right] = 0
\end{equation}
where the solution is 
\begin{equation}
K (n+1) \rho^{1/n} +\Psi_{\rm eff} =\Psi_{\rm eff,0}
\end{equation}
and $\Psi_{\rm eff,0}$ is the effective potential where the density, $\rho$ goes to zero. Finally, we can write an analytic expression for the density everywhere within star 1 as long as the effective potential is known:
\begin{equation}
\rho = \left[ \frac{\Psi_{\rm eff,0}-\Psi_{\rm eff}}{K(n+1)}\right]^n \, ,
\label{eqn:rho}
\end{equation}
see Eqn. (10) of \cite{hachisu1986}. In the case of a Roche-lobe filling star, $\Psi_{\rm eff,0} = \Psi_{\rm RL}$, where $\Psi_{\rm RL}$ is the potential at the Roche surface.

In order to solve for the density in Eqn.~(\ref{eqn:rho}) we need only approximate the potential of star 1 ($\Psi_1$), since $\Psi_2$ and $\Psi_{\rm cent}$ are already known.  Deep in the interior of star 1, the star is essentially spherically symmetric and largely unaffected by $\Psi_2$ and $\Psi_{\rm cent}$.  Thus we start the interior of star 1 by building a polytropic model.  Near where the surface of the unperturbed polytrope would be (and, in fact, well below its surface), the potential goes approximately as $\propto 1/r$ since most of the mass is concentrated near the center.  

The procedure we use is to construct a polytrope for the unperturbed problem with a radius that fits completely inside the equivalent Roche surface.  To this end, we define four distances from the center of star 1 to the unperturbed Roche potential, $R_{\rm x1}$, $R_{\rm x2}$, $R_{\rm y}$, $R_{\rm z}$.  These are along the $x$, $y$, and $z$ directions, as implied by the name, and the subscripts `1' and `2' refer to the direction toward the L$_1$ point and away from it, respectively.   We then set the nominal radius of the polytropic model to be $R_{\rm z}$, which is the smallest of the four distances.  Since HD~74423 is fairly massive ($M_1 \simeq 2.1 \, {\rm M}_\odot$) and hot ($T_{\rm eff} \simeq 7900$ K), we model it as an $n = 3$ polytrope, which does not yield an accurate density profile near the center (since the star is somewhat evolved), but should be adequate further out where we care most about the run of density and temperature.  Thus, from hereon, the discussion is not general for all polytropes and is limited to  $n=3$ polytropes.

The values of the polytropic constants for the problem, in terms of the mass, $M_1$, and radius, $R_1$ of the star are:
\begin{eqnarray}
K & =&  \pi G \left[\frac{M_1}{-4 \pi \xi_1^2 (d\phi/d\xi)_1} \right]^{2/3} \\
\rho_c & = & -\frac{\xi_1 M_1}{4 \pi R^3 (d\phi/d\xi)_1} \\
a & = & \sqrt{\frac{K}{\pi G}} ~\rho_c^{-1/3}
\end{eqnarray}
where $\rho_c$ is the central density, $a$ the length scale, $\phi$ the Lane-Emden solution for an $n=3$ polytrope, $\xi$ the dimensionless radial distance, $\xi_1$ is $\xi$ evaluated at the unperturbed surface of the polytrope, and $(d\phi/d\xi)_1$ is the derivative of the Lane-Emden function at the surface.  The density, temperature, and potential ($\Psi_1$) inside the star are:
\begin{eqnarray}
\rho(\xi) & = & \rho_c \phi(\xi)^n \\
T(\xi) & = & \frac{\mu {\rm m_p} K}{k} \rho_c^{1/n} \phi(\xi) \\
\Psi_1(\xi) & = & -\frac{GM_1}{R_1} - 4 K \rho_c^{1/3} \phi(\xi) 
\label{eqn:Psi}
\end{eqnarray}
where $k$ is Boltzmann's constant, $\mu {\rm m_p}$ is the mean molecular weight, and the radial distance $r \equiv a \xi$.

In order to make the model completely analytic, we utilize an approximation to the solution for the Lane-Emden equation for an $n=3$ polytrope:
\begin{equation}
\phi(\xi) \simeq \frac{1- (1/108)\xi^2 - (11/45 360)\xi^4}{1+(17/108)\xi^2 +(1/1008)\xi^4}
\end{equation}
where $\phi \rightarrow 0$ when $\xi = 6.89685$ for an $n=3$ polytrope.  This expression and its derivative (needed to compute the local gravity) are good to $\sim$1\%, which is adequate for our purposes.  

Thus, in Eq.~(\ref{eqn:rho}) we use $\Psi_1(\xi)$ from Eqn.~(\ref{eqn:Psi}) for $r = a \xi \lesssim R_z$, and $\Psi_1(r) = -GM_1/r$ for $r \gtrsim R_z$.  In fact, we use a hyperbolic tangent blending function between the two forms for $\Psi_1$ over a blending range of $\sim$10\% of $R_z$ to ensure a smooth transition. The density inside the Roche lobe is then everywhere determined analytically by the use of Eqn.~(\ref{eqn:rho}).  The pressure, temperature, and local sound speed follow from the polytropic relations, and $\vec{g}_{\rm eff}$ is found from Eqn.~(\ref{eqn:grav}).

Finally, in regard to the binary model, we note that the mass within the Roche lobe, but beyond a radial distance of $R_z$, contributes less than $\sim$0.1\% of the total mass $M_1$.  Therefore, the basic relation among the orbital period, $M_1$, $M_2$, and semi-major axis (i.e., Kepler's 3rd law), is not materially affected beyond the fractional percent level.

\bsp	
\label{lastpage}
\end{document}